\input harvmac

\Title{\vbox{\hbox{HUTP-97/A009}
\hbox{\tt hep-th/9702201}}}
{Lectures on Strings and Dualities}
\bigskip
\centerline{Cumrun Vafa}
\bigskip\centerline{\it Lyman Laboratory of Physics}
\centerline{\it Harvard University}\centerline{\it Cambridge, MA 02138}

\vskip .3in
In this set of lectures I review recent developments
in string theory emphasizing
their non-perturbative aspects and their recently discovered
duality symmetries.  The goal of the lectures is to make the recent exciting
developments in string theory accessible to those with
no previous background in string theory who wish to join the research
effort in this area.   Topics covered include
a brief review of string theory, its compactifications,
solitons and D-branes, black hole entropy and web of string dualities.
(Lectures presented at ICTP summer school, June 1996)
\Date{\it {February 1997}}

In this series of lectures I attempt to give
an overview of a subject roughly characterized as string duality.
String theory is by now a vast subject with almost three decades
of active research contributing to its development.  The past couple
of years have seen a tremendous growth in our understanding of string
theory, due to a better understanding of its non-perturbative
properties.  These developments have been so rapid that for a beginner
it may appear an impossible dream to join active research in this
exciting field of physics.  Here I hope to make a modest
effort towards bringing these highly advanced aspects of string theory
to a wider audience of physicits which have
no previous background in string theory.
Perhaps this will make
the beginners feel less shy about joining this
very exciting area of fundamental physics.

It is rather difficult (and probably impossible) to give
an overview of string theory and all the recent developments
in the relatively short space (and time!) available here.
My aim has been to present here some of the major highlights
with as much background as is needed to appreciate
their importance.  I hope that the interested reader will
continue by reading more in depth discussions
available in the literature.
Luckily there have been many excellent recent review
articles in this field \ref\review{
J. Schwarz,``Lectures on Superstring and M Theory Dualities,"
hep-th/9607201\semi
J. Polchinski,``String Duality--A Colloquium," hep-th/9607050\semi
J. Polchinski,``TASI Lectures on D-Branes,'' hep-th/9611050\semi
A. Sen,``Unification of String Dualities," hep-th/9609176 \semi
P. Townsend,``Four Lectures on M-theory," hep-th/9612121\semi
H. Ooguri and Z. Yin,``TASI Lectures on Perturbative String Theories,"
hep-th/9612254\semi
M. Duff,``Supermembranes," hep-th/9611203 \semi
M. Douglas,``Superstring Dualities, Dirichlet Branes and the Small Scale
Structure of Space," hep-th/9610041\semi
W. Lerche,``Introduction to Seiberg-Witten Theory and its Stringy Origin,"
hep-th/9611190\semi
P. Aspinwall,``K3 Surfaces and String Duality," hep-th/9611137\semi
G. Horowitz,``The Origin of Black Hole Entropy in String Theory,"
gr-qc/9604051\semi
B. Greene,``String Theory on Calabi-Yau Manifolds," hep-th/9702155.}\
to which I refer the interested reader to
for different emphasis on different aspects of string theory dualities.

The organization of this paper is as follows:  In section one
I make a lightning review of string theories.  In section two
I discuss compactification of string theories
(and in particular on tori, $K3$ manifold
and Calabi-Yau threefolds).  In section three I discuss aspects of
solitonic states in string theory.  In section four I discuss certain
implications of the existence of solitons for non-perturbative
aspects of string theory and entropy of extremal black holes.  In section
five I discuss the web of string dualities.  In section six I end with some
speculations and open questions.

I have decided to put no references in the main text (in the spirit
of giving an `informal' talk).  Apart from the above mentioned
review articles and references therein, let
me list some sample references for each section covered which
by no means is complete nor necessarily
the original works.  For sections one and two a related reference is
\ref\gsw{M. Green, J.H. Schwarz and E. Witten, ``Superstring Theory,"
(2 volumes), Cambridge University Press (1986).}\ref\oy{
H. Ooguri and Z. Yin,``TASI Lectures on Perturbative String Theories,"
hep-th/9612254.}\ and for $K3$ compactifications \ref\asp{P. Aspinwall,``K3
Surfaces and String Duality," hep-th/9611137.}.
  For section three see, for example, \ref\pol{J. Polchinski,``TASI Lectures on
D-Branes,'' hep-th/9611050}\ref\mdou{M. Douglas,``Superstring Dualities,
Dirichlet Branes and the Small Scale Structure of Space,"
hep-th/9610041.}\ref\St{A. Strominger,``Massless Black Holes and Conifolds
in String Theory," hep-th/9504090\semi B.Greene, D. Morrison and A. Strominger,
``Black Hole Condensation and the Unification of String Vacua,"
 hep-th/9504145.}.  For section four see, for example,
\ref\witb{E. Witten,``Bound States Of Strings And $p$-Branes,"
Nucl. Phys. B460 (1996) 335, hep-th/9510135.
}\ref\bsv{M. Bershadsky, V. Sadov and C. Vafa,``D-Strings
on D-manifolds," Nucl. Phys B463 (1996) 398, hep-th/
\semi M. Bershadsky, V. Sadov and C. Vafa,``D-branes and
topological field theories," Nucl. Phys. B463 (1996) 420,
hep-th/.}\ref\doet{M. Douglas, D. Kabat, P. Pouliot and S. Shenker,
``D-branes and Short Distances in String Theory,"
hep-th/9608024.}\ref\dvv{R. Dijkgraaf, E. Verlinde and H. Verlinde,
``BPS Quantization of the Five-Brane,"  hep-th/9604055.}\ and for
stringy aspects
of extremal black holes see \ref\sv{A. Strominger and C. Vafa,
``Microscopic origin of the Bekenstein-Hawking entropy,''
Phys. Lett. B383 (1996) 44,
hep-th 9601029}\ref\gho{G. Horowitz,``The Origin of Black Hole Entropy in
String Theory,"
gr-qc/9604051}.  For section five see
\ref\hult{C. Hull and P. Townsend,``Unity of Superstring Dualities,"
Nucl. Phys. B438 (1995) 109, hep-th/9410167}\ref\wit{E. Witten,``String Theory
Dynamics in Various Dimensions,"
Nucl. Phys. B443 (1995) 85, hep-th/9503124.}\ref\kv{S. Kachru and C.
Vafa,``Exact Results for N=2 Compactifications
of Heterotic Strings," Nucl. Phys. B450 (1995) 69, hep-th/9505105.}\ref\hwi{P.
Horava and E. Witten,``Heterotic and Type I String Dynamics from Eleven
Dimensions," hep-th/9510209.}\ref\pw{J. Polchinski and E. Witten,
``Evidence for Heterotic - Type I String Duality," Nucl. Phys. B460 (1996) 525,
hep-th/9510169.
}\ref\vf{C. Vafa, ``Evidence for F-theory," Nucl. Phys B469 (1996) 403,
hep-th/9602022.}\ref\asen{A. Sen,``Unification of String Dualities,"
hep-th/9609176.}.

\newsec{ Lightning Review of String Theory}

In this section we will give a brief overview of
string theory.  The interested reader is strongly advised
to study this subject more thorougly than presented here.
String theory is a description of dynamics of objects with
one spatial direction, which we parameterize by $\sigma$,
propagating in a space parameterized
by $X^\mu$.  The
 worldsheet of the string is
parameterized by coordinates
($\sigma ,\tau$) where each $\tau ={\rm
constant}$ denotes the string at a given time.
The amplitude for propagation of a string from an initial
configuration to a final one is given by sum over worldsheets which
interpolate between the two string configurations weighed by ${\rm exp}
(i S)$
where
\eqn\acti{S=\int d\sigma d\tau \ \partial_i X^\mu \partial_i X^{\nu}G_{\mu
\nu}(X)}
where $G_{\mu \nu}$ is the metric on spacetime
and $i$ runs over the $\sigma$ and $\tau $ directions.  Note that by
slicing the worldsheet we will get configurations where
a single string splits to a pair or vice versa, and combinations
thereof (consider for example the worldsheet configuration which
looks like a ``pant").

If we consider propagation in flat spacetime where $G_{\mu \nu}=\eta_{\mu \nu}$
the fields $X^{\mu}$
on the worldsheet, which describe the position in spacetime of each
bit of string, are free fields and satisfy the 2d equation
$$\partial _i\partial^i X^\mu =(\partial^2_\tau -\partial^2_\sigma )X^\mu =0$$
The solution of which is given by
$$X^\mu(\sigma, \tau)= X^\mu_L(\tau+\sigma )+X^\mu_R(\tau -\sigma )$$
In particular notice that the left- and right-moving degrees
of freedom are essentially independent.
There are two basic
types of strings:  {\it Closed} strings and {\it Open}
 strings depending on whether the string is a closed circle or an
open interval respectively.  If we are dealing with closed strings the
left- and right-moving degrees of freedom remain essentially independent
but if are dealing with open strings the left-moving modes
reflecting off the left boundary become the right-moving
modes--thus the left- and right-moving modes are essentially identical
in this case.  In this sense an open string has `half' the degrees
of freedom of a closed string and can be viewed as a `folding'
of a closed string so that it looks like an interval.

There are two basic types of string theories, bosonic and
fermionic.
What distinguishes bosonic and fermionic strings
is the existence of supersymmetry on the worldsheet.  This means
that in addition to the coordinates $X^\mu$ we also have anti-commuting
fermionic coordinates $\psi^{\mu}_{L,R}$ which are spacetime vectors but
fermionic
spinors on the worldsheet whose chirality is denoted by subscript $L,R$.  The
action for superstrings takes the form
$$S=\int \partial_LX^\mu \partial_R X^\mu +\psi_R^\mu \partial_L \psi_R^\mu
+\psi_L^\mu \partial_R \psi _R^\mu.$$
There are two consistent boundary conditions on each of the fermions, periodic
({\bf R}amond sector) or anti-periodic ({\bf N}eveu-{\bf S}chwarz sector)
(note that the coordinate $\sigma$ is periodic).

A natural question arises as to what metric
we should put on the worldsheet.  In the above
we have taken it to be flat.  However in principle there
is one degree of freedom that a metric can have in two dimensions.
This is because it is a $2\times 2$ symmetric matrix (3 degrees
of freedom) which is defined up to arbitrary reparametrization
of 2d spacetime (2 degrees of freedom) leaving us with one function.
Locally we can take the 2d metric $g$ to be conformally flat
$$g_{ij}={\rm exp}(\phi) \eta_{ij}$$
Classically the action $S$ does not depend on $\phi$.  This is easily seen by
noting that the properly coordinate invariant action density goes as
 ${\sqrt{g}}g^{ij}\partial_iX \partial_j X$ and is
independent of $\phi$ only in $d=2$.  This is rather nice and means that
we can ignore all the local dynamics associated with gravity on the
worldsheet.  This case is what is known as the critical string case
which is the case of most interest.  It turns out that this
independence from the local dynamics of the worldsheet metric
survives quantum corrections only when the dimension
of space is 26 in the case of {\it bosonic strings} and 10 for {\it fermionic}
or {\it superstrings}.  Each string can be in a specific vibrational mode which
gives
rise to a particle.  To describe the totality of such particles it is
convenient to go to `light-cone' gauge.  Roughly speaking this means
that we take into account that string vibration along their
worldsheet is not physical.  In particular for bosonic
string the vibrational modes exist only in 24 transverse directions
 and for superstrings they exist in $8$ transverse directions.

Solving the free field equations for $X,\psi $ we have
$$\partial_LX^\mu =
\sum_{n}\alpha_{-n}^\mu {\rm e}^{-in(\tau+\sigma)}$$
$$\psi_L^\mu =\sum_n \psi_{-n}^\mu {\rm e}^{-in(\tau+\sigma)}$$
and similarly for right-moving oscillator modes $\tilde \alpha_{-n}^\mu$
and $\tilde \psi_{-n}^\mu$.
The sum over $n$ in the above
runs over integers for the $\alpha_{-n}$.  For fermions
depending on whether we are in the {\bf{R}} sector or
{\bf NS} sector it runs over integers or integers shifted by ${1\over 2}$
respectively.
Many things decouple between the left- and right-movers in
the construction
of a single string Hilbert space and we sometimes talk only about
one of them.  For the open string Fock space the left- and right-movers
mix as mentioned before, and we simply get one copy of the above oscillators.

A special role is played by the zero modes
of the oscillators. For the $X$-fields they correspond
to the center of mass motion and thus $\alpha_0$ gets
identified with the left-moving momentum of the center of mass.
In particular we have for the center of mass
$$X=\alpha_0 (\tau+\sigma) +\tilde \alpha_0 (\tau -\sigma )$$
where we identify
$$(\alpha_0,\tilde \alpha_0)=(P_L,P_R)$$
Note that for closed string, periodicity of $X$ in $\sigma$ requires
that $P_L=P_R=P$ which we identify with the center of mass
momentum of the string.

In quantizing the fields on the strings we use the usual commutation
relations
$$[\alpha^\mu_n,\alpha^\nu_{m}]_-=n\delta_{m+n,0}\eta^{\mu \nu}$$
$$[\psi^\mu_n,\psi^\nu_{m}]_+=\eta^{\mu \nu}\delta_{m+n,0}$$
We choose the negative moded oscillators as creation operators.
In constructing the Fock space we have to pay special
attention to the zero modes.
The zero modes of $\alpha$ should be diagonal in
the Fock space and we identify
their eigenvalue with momentum.  For $\psi$ in the $NS$ sector
there is no zero mode so there is no subtlety in construction of
the Hilbert space.
For the $R$ sector, we have zero modes.  In this case the zero
modes form a Clifford algebra
$$[\psi^\mu_0,\psi^\nu_0]_+=\eta^{\mu \nu}.$$
This implies that in these cases the ground state is a spinor
representation of the Lorentz group.  Thus a typical element in the
fock space looks like
$$\alpha_{-n_1}^{L\mu_1}....\psi_{-n_k}^{L\mu_k}...|P_L,a\rangle\otimes
\alpha_{-m_1}^{R\mu_1}....\psi_{-m_r}^{R\mu_k}...|P_R,b\rangle $$
where $a,b$ label spinor states  for R sectors and are absent
in the NS case; moreover for the bosonic string we only have
the left and right bosonic oscillators.

It is convenient to define the total oscillator number as sum
of the negative oscillator numbers, for left- and right- movers separately.
$N_L=n_1+...+n_k+...$, $N_R=m_1+...+m_r+...$.
The condition that the two dimensional gravity decouple implies
that the energy momentum tensor annihilate the physical states.  The
trace of the energy momentum tensor is zero here (and in all
compactifications of string theory) and so we have two independent
components which can be identified with the left- and right-
moving hamiltonians $H_{L,R}$ and the physical states condition
requires that
\eqn\onsh{H_L=N_L+{1\over 2}P_L^2-\delta_L=0= H_R=N_R+{1\over
2}P_R^2-\delta_R}
where $\delta_{L,R}$ are normal ordering
constants which depend on which string theory and which sector
we are dealing with.  For bosonic string
$\delta =1$, for superstrings we have two cases:  For $NS$ sector
$\delta ={1\over 2}$ and for the $R$ sector $\delta =0$.
The equations \onsh\ give the spectrum of particles in the
string perturbation theory.  Note that $P_L^2=P_R^2=-m^2$
and so we see that $m^2$ grows linearly with the oscillator
number $N$, up to a shift:
\eqn\spec{{1\over 2}m^2=N_L-\delta_L=N_R-\delta_R}
 The number of states at oscillator number $N$ grows as
${\rm exp} \sqrt{cN}$  for some $c$, due to a large
number of ways we can get states at level $N$ by using oscillators.
Since $m\propto \sqrt N$ we learn that the number of states for large
mass $m$ grows as ${\rm exp}(\beta_H m)$ (for some constant $\beta_H$).
We thus have a
Hagedorn behaviour and a phase transition is expected at temperature
$T=1/\beta_H$.

Let us analyze the low lying states for bosoinc and
superstrings.
\subsec{Low Lying States of Bosonic Strings}

Let us consider the left-mover excitations.  Since
$\delta =1$ for bosonic string, \spec\ implies
that if we do not use any string oscillations, the
ground state is tachyonic ${1\over 2}m^2=-1$.  This clearly
implies that bosonic string by itself is not a good
starting point for perturbation theory.  Nevertheless
in anticipation of a modified appearance of bosonic
strings in the context of heterotic strings, let us
continue to the next state.

If we consider oscillator number $N_L=1$,
from \spec\ we learn that excitation is massless.  Putting the right-movers
together with it, we find that it is given by
$$\alpha^\mu_{-1}\tilde \alpha^\nu_{-1}|P\rangle$$
What is the physical interpretation of these massless states?
The most reliable method is to find how they transform under
the little group for massless states which in this case is
$SO(24)$.  If we go to the light cone gauge,
and count the physical states, which roughly
speaking means taking the indices $\mu$ to go over spatial
directions transverse to a null vector, we can easily deduce
the content of states.   By decomposing the above massless
state under the little group of $SO(24)$, we find that we
have symmetric traceless tensor, anti-symmetric 2-tensor, and the
trace, which we identify as arising from 26 dimensional fields
\eqn\grm{g_{\mu \nu},B_{\mu\nu},\phi}
the metric, the anti-symmetric field $B$ and the {\it dilaton}.
This triple of fields should be viewed as the stringy multiplet
for gravity.  The quantity $\lambda ={\rm \exp}[-\phi]$ is
identified with the the string coupling constant.  What this means
is that a worldsheet configuration of a string which sweeps a
genus $g$ curve, which should be viewed as $g$-th loop correction
for string theory, will be weighed by ${\rm
exp}(-2(g-1)\phi)=\lambda^{2g-2}$.
The existence of the field $B$ can also be understood (and
in some sense predicted) rather easily.
If we have a point particle it is natural to have it
charged under a gauge field, which introduces a
term ${\rm exp}(i\int A)$ along the worldline.  For strings
the natural generalization of this requires an anti-symmetric
2-form to integrate over the worldsheet, and so we say that
the strings are {\it charged} under $B_{\mu \nu}$ and that
the amplitude for a worldseet configuration will have
an extra factor of ${\rm exp}(i\int B)$.

Since bosonic string has tachyons we do not know how
to make sense of that theory by itself.

\subsec{Low Lying States of Type II Superstrings}
Let us now consider the light particle states for superstrings.
We recall from the above discussion that there are two
sectors to consider, NS and R, separately for the left- and the
right-movers.  As usual we will first treat the left- and right-moving
sectors separately and then combine them at the end.
Let us  consider
the NS sector for left-movers.
Then the formula for masses \onsh\ implies that the ground
state is tachyonic with ${1\over 2}m^2={-1\over 2}$.
The first excited states from the left-mover are massless
and corresponds to $\psi_{-1/2}^\mu |0\rangle$, and so
is a vector in spacetime.  How do we deal with the tachyons?
It turns out that summing
over the boundary conditions of fermions on the worldsheet amounts
to keeping the states with a fixed fermion number $(-1)^F$
on the worldsheet.
Since in the NS sector the number of fermionic oscillator
correlates with the integrality/half-integrality of $N$,
it turns out that the consistent choice involves keeping
only the $N=$half-integral states.  This is known
as the GSO projection.  Thus the tachyon is projected
out and the lightest left-moving state is a massless vector.

For the R-sector using \onsh\ we see that the ground
states are massless.  As discussed above, quantizing the
zero modes of fermions implies that they are spinors.  Moreover
GSO projection, which is projection on a definite $(-1)^F$ state,
amounts to projecting to spinors of a given chirality. So after
GSO projection we get a massless spinor of a definite chirality.
Let us denote the spinor of one chirality by $s$ and the other
one by $s'$.

Now let us combine the left- and right-moving sectors together.
Here we run into two distinct possibilities: A) The GSO
projections on the left- and right-movers are different
and lead in the R sector to ground states with different
chirality.
B) The GSO
projections on the left- and right-movers are the same
and lead in the R sector to ground states with the same
chirality.  The first case is known as type IIA superstring
and the second one as type IIB.  Let us see what kind
of massless modes we get for either of them.  From NS$\otimes$NS
we find for both type IIA,B
$$NS\otimes NS \rightarrow v\otimes v \rightarrow (g_{\mu \nu},B_{\mu
,\nu},\phi )$$
{}From the $NS \otimes R$ and $R \otimes NS$ we get the fermions
of the theory (including the gravitinos).  However the IIA and
IIB
differ in that the gravitinos of IIB are of the same chirality, whereas for
IIA they are of the opposite chirality.  This implies that IIB
is a chiral theory whereas IIA is non-chiral.
Let us move to the R$\otimes$ R sector.  We find
$$IIA: R\otimes R =s\otimes s' \rightarrow (A_\mu, C_{\mu \nu \rho})$$
\eqn\cont{IIB: R\otimes R =s\otimes s\rightarrow (\chi,B'_{\mu \nu},
D_{\mu\nu\rho \lambda})}
where all the tensors appearing above are fully antisymmetric.  Moreover
$D_{\mu \nu \rho \lambda}$ has a self-dual field strength $F=dD=*F$.
It turns out that to write the equations of motion in a unified way
it is convenient to consider a generalized gauge fields ${\cal A}$
and ${\cal B}$ in the IIA and IIB case respectively by adding all
the fields in the RR sector together
with the following
properties: i)
${\cal A} ({\cal B})$ involve all the odd (even) dimensional
antisymmetric fields. ii) the equation of motion is
$d{\cal A}=* d{\cal A}$.  In the case of all fields
(except $D_{\mu\nu\lambda \rho}$) this equation allows
us to solve for the forms with degrees bigger than 4 in terms
of the lower ones and moreover it implies the field equation
$d*dA=0$ which is the familiar field equation for the guage fields.
In the case of the $D$-field it simply gives that its field strength
is self-dual.

\subsec{Open Superstring: Type I string}

In the case of type IIB theory in 10 dimensions, we note
that the left- and right-moving degrees of freedom on the worldsheet
are the same.  In this case we can `mod out' by a reflection symmetry
on the string; this means keeping only the states in the full Hilbert
space which are invariant under the left-/right-moving exchange of quantum
numbers.  This is simply projecting the hilbert space onto the
invariant subspace of the projection operator $P={1\over 2}(1+\Omega)$ where
$\Omega$ exchanges left- and right-movers.   $\Omega$ is known
as the orientifold operation as it reverses the orientation
on the worldsheet.
 Note that this symmetry
only exists for IIB and not for IIA theory (unless
we accompany it with a parity reflection in spacetime).  Let us see which
bosonic states we will be left with after this projection.
{}From the NS-NS sector $B_{\mu\nu}$ is odd and projected
out and thus we are left with the symmetric parts of the tensor product
$$NS-NS\rightarrow (v\otimes v)_{symm.}=(g_{\mu \nu},\phi )$$
{}From the R-R sector since the degrees of freedom are fermionic
from each sector we get, when exchanging left- and right-movers
an extra minus sign which thus means we have to keep anti-symmetric
parts of the tensor product
$$R-R\rightarrow (s\otimes s)_{anti-symm.}={\tilde B}_{\mu \nu}$$
This is not the end of the story, however.  In order to make
the theory consistent we need to introduce a new sector
in this theory involving open strings.  This comes about
from the fact that in the R-R sector there actually is
a 10 form gauge potential which has no propagating
degree of freedom, but acquires a tadpole.  Introduction
of a suitable open string sector cancels this tadpole.

As noted before the construction of open string sector Hilbert space proceeds
as in the closed string case, but now, the left-moving and
right-moving modes become indistinguishable due to reflection
off the boundaries of open string.  We thus get only one copy
of the oscillators. Moreover we can associate `Chan-Paton' factors
to the boundaries of open string (very much as in the picture for
mesons made of quark-antiquark pairs connected by flux lines).
To cancel the tadpole it turns out that we need 32 Chan-Paton
labels on each end.  We still have two sectors corresponding
to the NS and R sectors.  The NS sector gives a vector field
$A_\mu$ and the R sector gives the gaugino.  The gauge
field $A_{\mu}$ has two additional labels coming
from the end points of the open string and it turns out
that the left-right exchange projection of the type IIB theory
translates to keeping the antisymmetric component
of $A_\mu =-A_\mu^T$, which means we have an adjoint of $SO(32)$.
Thus all put together, the bosonic
degrees of freedom are
$$(g_{\mu \nu},{\tilde B}_{\mu\nu},\phi)+(A_\mu)_{SO(32)}$$
We should keep in mind here that $\tilde B$ came not from the
NS-NS sector, but from the R-R sector.

\subsec{Heterotic Strings}
Heterotic string is a combination of bosonic string and
superstring, where roughly speaking the left-moving
degrees of freedom are as in the bosonic string and
the right-moving degrees of freedom are as in the superstring.
It is clear that this makes sense for the construction of the
states because the left- and right-moving sectors hardly
talk with each other.  This is almost true, however they
are linked together by the zero modes of the bosonic oscillators
which give rise to momenta $(P_L,P_R)$.  Previously we had
$P_L=P_R$ but now this cannot be the case because $P_L$
is 26 dimensional but $P_R$ is 10 dimensional.  It is
natural to decompose $P_L$ to a 10+16 dimensional vectors,
where we identify the 10 dimensional part of it with $P_R$.
It turns out that for the consistency of the theory the extra
16 dimensional component should belong to the root lattice
of $E_8\times E_8$ or a $Z_2$ sublattice of $SO(32)$ weight lattice.
In either of these two cases the vectors in the lattice with $(length)^2=2$
are in one to one correspondence with non-zero weights in the adjoint
of $E_8\times E_8$ and $SO(32)$ respectively.  These can
also be conveniently represented (through bosonization)
by 32 fermions:  In the case of $E_8\times E_8$ we group
them to two groups of 16 and consider independent NS, R
sectors for each group.
 In the case of $SO(32)$ we only
have one group of $32$ fermions with either NS or R boundary
conditions.

Let us tabulate the massless modes using \spec .
The right-movers can be either NS or R.
The left-moving degrees of freedom start out with a tachyonic
mode.  But \spec\ implies that this is not satisfying
the level-matching condition because the right-moving ground state is at
zero energy.  Thus we should search on the left-moving side
for states with $L_0=0$ which means from \spec\ that we have
either $N_L=1$ or ${1\over 2}P_L^2=1$, where $P_L$ is an internal
16 dimensional vector in one of the two lattices noted above.
The states with $N_L=1$ are
$$16\oplus v$$
where 16 corresponds to the oscillation direction in the
extra 16 dimensions and $v$ corresponds to vector
in 10 dimensional spacetime.  States with ${1\over 2}P_L^2=1$
correspond to the non-zero weights of the adjoint of $E_8\times E_8$
or $SO(32)$ which altogether correspond to $480$ states in both
cases.  The extra 16 $N_L=1$ modes combine with these $480$ states
to form the adjoints of $E_8\times E_8$ or $SO(32)$ respectively.
The right-movers give, as before, a $v\oplus s$ from the NS and R
sectors respectively.  So putting the left- and right-movers
together we finally get for the massless modes
$$(v\oplus Adj)\otimes (v\oplus s)$$
Thus the bosonic states are $(v\oplus Adj)\otimes v$ which gives
$$(g_{\mu \nu},B_{\mu \nu},\phi ; A_{\mu})$$
where the $A_\mu$ is in the adjoint of $E_8\times E_8$ or
$SO(32)$.  Note that in the $SO(32)$ case this is an
{\it identical} spectrum to that of type I strings.

\subsec{Summary}
Just to summarize, we have found 5 consistent strings in 10
dimensions:  Type IIA with $N=2$ non-chiral supersymmetry, type IIB
with $N=2$ chiral supersymmetry, type I with N=1 supersymmetry and
gauge symmetry $SO(32)$ and heterotic strings with N=1
supersymmetry with $SO(32)$ or $E_8\times E_8$ gauge symmetry.
Note that as far as the massless modes are concerned
we only have four inequivalent theories, because heterotic
$SO(32)$ theory and Type I theory have the same light degrees of freedom.
We shall return to this point later, when we discuss dualities.
In discussing compactifications it is sometimes natural to divide
the discussion between two cases depending on how many supersymmetries
we start with.  In this context we will refer to the type IIA and B
as $N=2$ {\it theories} and Type I and heterotic strings as $N=1$ {\it
theories}.

\newsec{String Compactifications}
So far we have only talked about superstrings propagating
in 10 dimensional Minkowski spacetime.  If we wish
to connect string theory to the observed four dimensional
spacetime, somehow we have to get rid of the extra 6 directions.
One way to do this is by assuming that the extra 6 dimensions
are tiny and thus unobservable in the present day experiments.
In such scenarios we have to understand strings propagating
not on ten dimensional Minkowski spacetime but on four dimensional
Minkowski spacetime times a compact 6 dimensional manifold $K$.  In order to
gain more insight it is convenient
to consider compactifications not just to 4 dimensions but
to arbitrary dimensional spacetimes, in which case the
dimension of $K$ is variable.

The choice of $K$ and the string theory we choose to start in 10 dimensions
will lead to a large number of
 theories in diverse dimensions,
which have different number of supersymmetries and different
low energy effective degrees of freedom. In order to get a handle on such
compactifications it is useful to first classify them according to
how much supersymmetry they preserve.  This is useful because the
higher the number of supersymmetry the less the quantum
corrections there are.

If we consider a general manifold $K$ we find that
the supersymmetry is completely broken.  This is the case
we would really like to understand, but it turns out that
string theory perturbation theory always breaks down
in such a situation;  this is intimately connected
with the fact that typically cosmological constant is
generated by perturbation theory and this destablizes the Minkowski
solution.   For this
reason we do not even have a single example of such a class whose dynamics
we understand.  Instead if we choose $K$ to be of a special
type we can preserve a number of supersymmetries.

For this to be the case, we need $K$ to admit some number
of covariantly constant spinors.  This is the case
because the number of supercharges which are `unbroken' by
compactification is related to how many covariantly constant
spinors we have. To see this note that if we wish to define
a {\it constant} supersymmetry transformation, since a spacetime
spinor, is also a spinor of internal space, we need in addition a
 constant spinor in the internal compact directions.
 The basic choices are manifolds with
trivial holonomy (flat tori are the only example), $SU(n)$ holonomy
(Calabi-Yau n-folds), $Sp(n)$ holonomy (4n dimensional manifolds),
 7-manifolds of $G_2$ holonomy
and 8-manifolds of $Spin(7)$ holonomy.  The case mostly studied
in physics involves toroidal compactification, $SU(2)=Sp(1)$ holonomy
manifold (the 4-dimensional $K3$), $SU(3)$ holonomy (Calabi-Yau
3-folds)\foot{Calabi-Yau 4-folds have also recently appeared
in connection with F-theory compactification to 4 dimensions
as we will briefly discuss later.}.
The cases of $Sp(2)$ holonomy manifolds (8 dimensional)
and $G_2$ and $Spin(7)$ end up giving us compactifications
below 4 dimensions.
Here we will review the cases of toroidal compactification,
$K3$ compactification and Calabi-Yau 3-fold compactification.

\subsec{Toroidal Compactifications}
The space with maximal number of covariantly constant
spinors is the flat torus $T^d$.  This is also the easiest
to describe the string propagation in.
The main modification to the construction of the Hilbert space
from flat non-compact space in this
case involves relaxing
the condition $P_L=P_R$ because the string
can wrap around the internal space and so $X$ does not need
to come back to itself as we go around $\sigma$ .  In particular if we
consider compactification on
 a circle of radius $R$
we can have
$$(P_L,P_R)=({n\over 2R}+mR,{n\over 2R}-mR)$$
Here $n$ labels the center of mass momentum of the string along
the circle and $m$ labels how many times the string is winding around
the circle.  Note that the spectrum of allowed $(P_L,P_R)$ is invariant
under $R\rightarrow 1/2R$.  All that we have to do is to exchange
the momentum and winding modes ($n\leftrightarrow m$).
This symmetry is a consequence of what is
known as $T$-duality and as you can see it is a relatively simple
fact to understand.
If we compactify on a d-dimensional torus $T^d$ it can be shown
that $(P_L,P_R)$ belong to a 2d dimensional lattice with
signature $(d,d)$.  Moreoever this lattice is integral, self-dual and even.
Evenness means, $P_L^2-P_R^2$ is even
for each lattice vector.  Self-duality means
that any vector which has integral product with all the vectors in the lattice
sits in the lattice as well.  It is an easy exercise to check these condition
in the one dimensional circle example given above.  Note
that we can change the radii of the torus and this will
clearly affect the $(P_L,P_R)$.  Given any choice of a
d-dimensional torus compactifications, all the other ones
can be obtained by doing an $SO(d,d)$ Lorentz boost on $(P_L,P_R)$ vectors.
Of course rotating $(P_L,P_R)$ by an $O(d)\times O(d)$ transformation
does not change the spectrum of the string states, so the
totality of such vectors is given by
$$SO(d,d)\over SO(d)\times SO(d)$$
Some Lorentz boosts will not change the lattice and amount to relabeling
the states.  These are the boosts that sit in $O(d,d;Z)$ (i.e. boosts
with integer coefficients), because they can be undone by choosing
a new basis for the lattice by taking
an integral linear combination of lattice vectors.  So the space of
inequivalent choices are actually given by
$$SO(d,d)\over SO(d)\times SO(d)\times O(d,d;Z)$$
The $O(d,d;Z)$ generalizes the T-duality considered in the
1-dimensional case.  So far our discussion is general in that
we have described only the bosonic degrees of freedom of the
string.  In discussing further aspects it is useful
to divide the discussion to the toroidal compactification
of $N=2$ supersymmetric theories (IIA and IIB) and
$N=1$ case (Type I and heterotic).

\subsec{$N=2$ theories on $T^d$}
Once we compactify type IIA and IIB on a circle
both theories have the same low energy degrees of freedom.
Actually they are isomorphic to each other; the T-duality
discussed before, $R\rightarrow {1\over R}$, exchanges
the two theories.  This is because the operation of $R\rightarrow
{1\over R}$ is accompanied on the fermions along the circle direction by
$$\psi_L \rightarrow -\psi_L$$
$$\psi_R \rightarrow \psi_R$$
This in particular means that the product of all the left
moving fermions, which defines the left-moving chirality,
is switched by an overall minus sign, and thus under the GSO
projection we keep the opposite chirality to what we
had.  This therefore exchanges $IIA$ and $IIB$.  For more general
compactification on $T^d$ the part of the T-duality group
which does not exchange the two theories is $SO(d,d;Z)$; the elements
of T-duality which are in $O(d,d;Z)$ but not in $SO(d,d;Z)$ will
exchange IIA and IIB and thus are not symmetries of either one.

In compactifying type II strings on tori, the scalars parameterized
by the coset $SO(d,d)/SO(d)\times SO(d)$ correspond
to choices of the metric of the torus  ($d(d+1)/2 $ degrees of freedom)
and the anti-symmetric field $B_{ij}$
on the torus ($d(d-1)/2$ degrees of freedom).  As we see
these do modify the string Hilbert space spectrum.  However there
are other choices we have to make by turning on expectation
values of the RR anti-symmetric tensor fields with zero field
strength.  Since none of the string modes couple
to the anti-symmetric potentials and only couple
to the field strength, this will not affect the perturbative
string spectrum.  It will however affect the solitonic
spectrum of string states.

Let us count how many total parameters
we have in specifying the choice of a vacuum.  These
would correspond to massless scalars in the non-compact theory.
Just as an example let us consider
the compactification of type IIA strings on $T^4$.  Then we have
$4\times 4=16$ parameters specifying the metric of $T^4$ as well
as the anti-symmetric field on it.  In addition we have 4 parameters
specifiying the choice for the Wilson line of $A_{\mu}$ and
4 parameters specifying the choice for the constant modes of
$C_{\mu\nu\rho}$.  So altogether we have $16+4+4=24$ massless
scalars coming from the choice
of the compactification.  Noting that we also have one scalar
corresponding to the dilaton in 10 dimensions we have altogether
25 massless scalars.
  The kinetic term for 16 of these scalars corresponding
to the internal metric and antisymmetric $B$-field is given
by $SO(4,4)/SO(4)\times SO(4)$ coset metric.  One could
ask about what the metric looks like for the other parameters?
It turns out that supersymmetry alone predicts that the metric
of the total 25 dimensional space is that of the coset
$$SO(5,5)/SO(5)\times SO(5)$$
This appears somewhat mysterious from the viewpoint
of string theory, because the origin of various scalars
are so different.  The above coset unifies dilaton,
RR tensor fields and the metric degrees of freedom of the
compactification into one single object!
The simplest way to understand this result
of supergravity is as follows:  The low energy
degrees of freedom, which is all that is needed
in the description of the coset, can
also be obtained from the compactification of $N=1$ supergravity
in eleven dimensions
on a 5-dimensional torus.  The choice of the metric on $T^5$ up
to an overall scale involves the coset space
$$SL(5)/SO(5)$$
The group $SL(5)$ and the group
visible from string perturbation theory, namely $SO(4,4)$
intersect on $SL(4)$.  The smallest group containing
both is $SO(5,5)$.  This gives the coset description
given above, a fact which also follows from supersymmetry arguments
alone.
Note that this same reasoning would lead
us to postulate the global identifications on this coset space
to be $SO(5,5;{\bf Z})$ because the identification
from string theory side is $SO(4,4;{\bf Z})$
and from the compactification from $N=1$ theory in eleven
dimensions $SL(5;{\bf Z})$ which together generate
$SO(5,5;{\bf Z})$.  Thus if we believe that the compactification
of the 11 dimensional theory is somehow equivalent to type II
strings, as we will discuss later, we would be led to
 the coset space
$$SO(5,5)/SO(5)\times SO(5)\times SO(5,5;Z).$$
  Note that
the elements in $SO(5,5;Z)$ not contained in $SO(4,4;Z)$ act
non-trivially on the dilaton of type IIA strings.  This
symmetry cannot be directly checked in string perturbation
theory because it typically takes small values of the string coupling
to large values, inaccessible by string perturbation techniques.
We will return to evidence for this kind of non-perturbative
symmetry, known as U-dualities later on in this paper.

The same construction generalizes to compactification on
$T^d$.  In this case the T-duality group of string theory
is $SO(d,d)$ and that of the 11 dimensional theory is $SL(d+1)$,
which have the common subgroup $SL(d)$.  Together
they form the group $E_{d+1}$ which is the maximally non-compact
form of the exceptional group.  Note that upon compactification
of all spatial directions we end up with $E_{10}$ which is the
hyperbolic Kac-Moody algebra.  This suggests that $E_{10}$
should play a prominent role in an ``invariant'' way of thinking
about type II string theories.

\subsec{Compactifications of $N=1$ theories on $T^d$}
If we compactify the heterotic strings ($E_8\times E_8$ or $SO(32)$)
or Type I theory on $T^d$, in addition to the choice of the
metric $g_{ij}$ of $T^d$ and the antisymmetric field $B_{ij}$
on it, whose moduli form the same coset as for the type II
case discussed above $SO(d,d)/SO(d)\times SO(d)$, we also have
to choose Wilson lines which lie in the Cartan of these gauge groups.
Given that in all these cases the rank of the group is $16$ the local
structure turns out to be the coset space
$${SO(d+16,d)\over SO(d+16)\times SO(d)}.$$
Note that the dimension of this space is $16 d$ bigger
than that for the type II case, which signifies the choice
of $16$ $U(1)$ Wilson lines along $d$ different directions.
For the global aspects of identification, the heterotic case
is very simple and leads to $SO(d+16,d;{\bf Z})$ symmetry in
the denominator above.  In fact the construction of Hilbert space
is particularly simple and the only difference from the 10 dimensional
case is that now the left and right momenta can lie on a self-dual
even lorentzian lattice of signature $(16+d,d)$, denoted by
$\Gamma^{16+d,d}$.  In other words
$$(P_L,P_R)\in \Gamma^{16+d,d}$$
It is known that this lattice is unique.  The only choice
comes from the `choice of polarization', which corresponds to how
we project to left- and right- momenta.  The moduli space
for this choice is precisely
$${SO(16+d,d;{\bf R})\over SO(d+16; {\bf R})\times SO(d:{\bf R})
\times SO(16+d,d;{\bf Z})}$$
This uniqueness implies in particular that upon toroidal
compactification (in fact even upon compactification on a circle)
the $E_8\times E_8$ heterotic string and $SO(32)$ heterotic string
become isomorphic, i.e. T-dual to one another.

For the type I string the local part of the moduli space
is easily seen to be the same and it is conjectured
to also have the same global T-duality group $SO(d+16,d:{\bf Z})$.

Note that apart from the above moduli, the expectation
value of the dilaton, $\phi$ which plays the role
of the string coupling constant, is part of the moduli
of vacua of this theory.  For $d\leq 4$ there are extra scalars.
Let us consider the case where $d=4$.  In this case we have
an $N=4$ supersymmetric theory.  In addition to the above moduli
the anti-symmetric
field in 4 dimension is dual to a scalar and so combining
this with the dilaton we have a complex field.  Moreover the $N=4$
supergravity fixes the metric on that space to be exactly
the same metric as the upper-half plane.  The moduli
space of this theory in $d=4$ is locally given by
$${SL(2;{\bf R})\over SO(2;{\bf R})}\times {SO(22,6; {\bf R})\over
SO(22;{\bf R})\times SO(6;{\bf R})}$$
It is natural to speculate that the global symmetry of the
first factor is $SL(2,{\bf Z})$.  The non-trivial
element of this duality acts to invert the four dimensional
string coupling constant and conjecturing such a symmetry
is consistent with the Olive-Montonen strong/weak duality
conjecture for $N=4$ field theories.  We shall return to this point
later.

\subsec{Compactifications on $K3$}
The four dimensional manifold $K3$ is the only manifold in four dimensions,
besides $T^4$, which admits covariantly constant spinors.  In fact
it has exactly half the number of covariantly constant spinors
as on $T^4$ and thus preserves half of the supersymmetry that would
have been preserved upon toroidal compactification. More precisely
the holonomy of a generic four manifold is $SO(4)$.  If the holonomy
resides in an $SU(2)$ subgroup of $SO(4)$ which leaves an $SU(2)$
part of $SO(4)$ untouched, we end up with one chirality of $SO(4)$
spinor being unaffected by the curvature of $K3$, which allows
us to define supersymmetry transformations as if $K3$ were flat
(note a spinor of $SO(4)$ decomposes as $({\bf 2},
{\bf 1})\oplus ({\bf 1},{\bf 2})$ of $SU(2)\times SU(2)$).
 Before considering
compactifications of various superstrings on $K3$ it is convenient to talk
about various geometric properties of $K3$.

\subsec{Aspects of $K3$ manifold}
There are a number of realizations of $K3$, which are useful
depending on which question one is interested in.
We shall describe all these different view points
in turn.  Perhaps
the simplest description of it is in terms of {\it orbifolds}.
This description of $K3$ is very close to toroidal compactification
and differs from it by certain discrete isometries of the $T^4$
which are used to (generically) identify points which are in
the same {\it orbit} of the discrete group. The other description
of $K3$ describes a 19 complex parameter family of $K3$ defined
by an algebraic equation.   Finally an 18 complex parameter
subspace of $K3$ surfaces admit elliptic fibrations (with
a section) which is an intermediate between the toroidal
orbifold limit of $K3$ and its algebraic description.

\subsec{$K3$ as an orbifold of $T^4$}

Consider a $T^4$ which for simplicity we take to be
parametrized by four real coordinates $x_i$ with $i=1,...,4$,
subject to the identifications $x_i\sim x_i+1$. It is sometimes
convenient to think of this as two complex coordinates $z_1=x_1+ix_2$
and $z_2=x_3+ix_4$ with the obvious identifications.  Now we identify
the points on the torus which are mapped to each other under the $Z_2$
action (involution) given by reflection in the coordinates
$x_i\rightarrow -x_i$, which is equivalent to
$$z_i\rightarrow -z_i$$
Note that this action has $2^4=16$ fixed points given by the choice
of midpoints or the origin in any of the four $x_i$.
The resulting space is singular at any of these 16 fixed points
because the angular degree of freedom around each of these
points is cut by half.  Put differently, if we consider any primitive
loop going `around' any of these 16 fixed point, it corresponds
to an open curve on $T^4$ which connects pairs of points related
by the $Z_2$ involution.  Moreover the parallel transport of vectors
along this path, after using the $Z_2$ identification, results
in a flip of the sign of the vector.  This is true no matter
how small the curve is.  This shows that we cannot have a smooth
manifold at the fixed points.

One of the nice things about the orbifold limit of $K3$ is that
the construction of the string Hilbert space is relatively simple
(known as the orbifold construction): we start with the Hilbert space
of $T^4$ compactification.  The $Z_2$ action on the torus will induce
a $Z_2$ action on the Hilbert space.
We project unto the $Z_2$ invariant
subspace. This is not the end of story for closed strings on $K3$:
We will have to consider open strings on $T^4$ which begin
and end at points identified under the $Z_2$ map. This is called
the twisted sector.  In this case we have 16 choices of the center
of mass of the twisted string state. Moreover the bosonic oscillators
are shifted by $1/2$ (because we have anti-periodic boundary
conditions induced by $x_i(0)=-x_i(2\pi)$).  Clearly these modifications
are rather minor and it is straightforward to consider
the twisted sector.  The only point is that now we have no zero modes
corresponding to left or right momenta.  In fact the center of mass
of the twisted string is frozen at any of the 16 fixed points of the
$Z_2$ action.  To complete the orbifold construction we still
have to project unto the $Z_2$ invariant subsector even in the
twisted sector.

As noted above this is a singular limit of $K3$, due to the singularities
of the metric at the 16 fixed points.  It is natural
to ask how these singularities can be remedied.

To answer the first question it is convenient to use complex coordinates.
Near a fixed point $(z_1,z_2)=(0,0)$, where the identification is
$(z_1,z_2)\rightarrow (-z_1,-z_2)$, it is natural to choose
coordinates that are invariant under this action.  To do this
we use the complex coordinates
$$u=z_1z_2$$
$$v=z_1^2$$
$$w=z_2^2$$
Clearly these are the only basic $Z_2$ invariant combinations
of these coordinates.  However there is a relation between them:
$$u^2=vw$$
Since manifolds can be identified by the functions on them,
the above equation can also be viewed as being equivalent to the
$K3$ manifold near the singular locus.
By a simple change of complex coordinates this can also
be written as
\eqn\sina{f=u^2+{\tilde v}^2+{\tilde w}^2=0}
It is easy to see that at $u={\tilde v}={\tilde w}=0$ the manifold
given by the above equation
is singular (by considering the rank of $df$ at that point).
This of course was expected as it is equivalent to the fixed point
of the $Z_2$ action discussed above.   This way of writing
the singularity also makes clear that there is a 2-sphere hidden
in the singularity.  Namely, if we consider the sublocus
of the manifold where $u,{\tilde v},{\tilde w}$ are
 real, then $f=0$ is the equation for a sphere of zero size.
In other words we have a {\it vanishing sphere} at this
singularity.

Now we come to how this singularity can be resolved.
In general this can be done in two ways.  Given
the discussion of the existence of the singularity at the
origin it is natural to eliminate that point from the
manifold by slightly changing the defining equation of the manifold.
  In other words we can {\it deform the complex
structure} of the manifold by considering
$$f=u^2+{\tilde v}^2+{\tilde w}^2=\epsilon$$
In this way the point $u={\tilde v}={\tilde w}=0$ is eliminated
from the manifold and in fact the manifold is now smooth.
Another way this singularity can be remedied is by giving the sphere
defined by $f=0$ a finite volume.  This is known as {\it blowing up}
the singularity.
Changing the volumes of the complex manifold
are known as {\it Kahler deformations}.  So in this
case we say that the singularity of the manifold has been
repaired by {\it deformation of the Kahler structure}.
    It turns out that in the special case
of $K3$ these two operations are not unrelated.  In fact if
we take $\epsilon$ to be real and view it as radius squared
of the sphere, the above equation can also be viewed as giving
the vanishing sphere a finite volume.
Only in the case of $K3$ the deformation by Kahler
and complex structure can be related (this has to do
with the fact that $K3$ is hyperkahler). In more general
cases, such as Calabi-Yau threefolds, the deformations
of Kahler and complex structures are distinguishable and repairing
singularities by using them in
general lead to topologically distinct manifolds.

The above singularity is known as an $A_1$ singularity.
There is a simple generalization of it which occurs when
we locally identify:
$$z_1\rightarrow \omega z_1$$
$$z_2\rightarrow \omega^{-1}z_2$$
where $\omega$ is an $n$-th root of unity.
This space is known as an ALE (asymptotically locally
Euclidean) space.  In this case we define
the invariant coordinates as
$$u=z_1z_2$$
$$v=z_1^n$$
$$w=z_2^n$$
with the identification
$$u^n=vw$$
which defines the local description of the singularity.  To
resolve the singularity we introduce terms of lower order in $u$
and consider
$$\prod_{i=1}^n (u-a_i)=vw=-{\tilde v}^2-{\tilde w}^2$$
If any of the two $a_i$ are equal then we have a singularity
of the type $A_1$ discussed before and we obtain
a vanishing 2-sphere.  When all $a_i$ are distinct
the singularity is completely resolved. In general
to any pair of $a_i$ we can associate a vanishing 2-sphere.
If we wish to express the homology class they represent  all of them
can be represented as a linear combination of
$n-1$ vanishing 2-spheres.   For example, we can take all $a_i$
to be real and order the index of $a_i$ to be in an increasing order.
We then take the $j$-th generator $S_j$ to be the vanishing
sphere associated with $a_j$ approaching $a_{j+1}$.
If we consider the image of $S_j$ on the $u$-plane it
corresponds to an interval running from $a_j$ to $a_{j+1}$.
In this way it is also easy to see that the $S_j$ and $S_{j+1}$
intersect one another at one point, namely $u=a_{j+1},{\tilde v}={\tilde w}=0$.
If for every generator $S_j$ we consider a node and for every
intersection between adjacent vanishing spheres draw a line
connecting them we obtain the $A_{n-1}$ Dynkin diagram.  Indeed
this singularity is known as $A_{n-1}$ singularity.
Note that again as in the $A_1$ case we can consider blowing up
the singularity instead of deforming its complex structure.  However
again in the case of $K3$ this turns out to be equivalent to deformation.

There are extensions of this singularity associated with modding out
the $(z_1,z_2)$ complex space by Dihedral and exceptional subgroups
of $SU(2)$.  In these cases the story is very similar
to what we discussed above.  When we consider the generators
for the vanishing spheres and their intersections we get
the Dynkin diagram for the D and E series (corresponding respectively
to the Dihedral and exceptional subgroups of $SU(2)$).

Before moving to other descriptions of $K3$ let us note
that the cohomology classes of $K3$ are very visible in
the orbifold construction.   We decompose the cohomology
classes according to the degree of holomorphic and
anti-holomorphic forms $(p,q)$ and denote the number
in each class by $h^{p,q}$
We get from the untwisted sector 8 elements corresponding
to the $Z_2$ invariant forms on $T^4$, namely,
 $$1,dz_1\wedge dz_2,
d{\overline z}_1\wedge d{\overline z}_2,dz_1\wedge dz_2\wedge
d{\overline z}_1\wedge d{\overline z}_2$$
$$dz_i\wedge d{\overline z}_j$$
{}From the twisted sectors, after resolving we obtain
from each a contribution to $h^{1,1}$.  This is the cohomology
vanishing 2-sphere hidden in the fixed points, discussed above.
So we obtain altogether 16 elements in $h^{1,1}$ coming from
the twisted sector.    Note that the number of Kahler deformations
is the dimension of $h^{1,1}$ which altogether is $20$.  Also
the number of complex deformations for $K3$ is also given by $h^{1,1}$
which is 20 (this will be made
more precise below); 16 of them correspond to resolving
the singularity near the fixed points and 4 of them correspond
to changing the complex structure of the underlying $T^4$.
Note that from the cohomology computation above
 we learn in particular that
the Euler characteristic of $K3$ is $\chi(K3)=24$.

\subsec{$K3$ as a complex surface in ${\bf P^3}$}
When we move away from the orbifold points of $K3$ the description
of the geometry of $K3$ in terms of the properties of the $T^4$
and the $Z_2$ twist become less relevant, and it is natural
to ask about other ways to think about $K3$.  In general
a simple way to define complex manifolds is by imposing
complex equations in a compact space known as
the projective $n$-space ${\bf P}^n$.  This is the space
of complex variables $(z_1,...,z_{n+1})$ excluding
the origin and subject to the identification
$$(z_1,...,z_{n+1})\sim \lambda (z_1,...,z_{n+1}) \qquad \lambda \not=0$$
One then considers the vanishing
locus of a homogeneous polynomial of degree $d$, $W_d(z_i)=0$ to obtain
an $n-1$ dimensional subspace of ${\bf P}^n$.
An interesting special case is when the degree is $d=n+1$.
In this case one obtains an $n-1$ complex dimensional manifold
which admits a Ricci-flat metric.  This is the case known as Calabi-Yau.
For example, if we take the case $n=2$, by considering
cubics in it
$$z_1^3+z_2^3+z_3^3+az_1z_2z_3=0$$
we obtain an elliptic curve, i.e. a torus of complex dimension
1 or real dimension 2.  The next case would be $n=3$ in which
case, if we consider a quartic polynomial in ${\bf P}^3$
we obtain the 2 complex dimensional $K3$ manifold:
$$W=z_1^4+z_2^4+z_3^4+z_4^4 +{\rm deformations}=0$$
There are 19 inequivalent quartic terms we can add.  This
gives us a 19 dimensional complex subspace of 20
dimensional complex moduli of the $K3$ manifold.  Clearly this
way of representing $K3$ makes the complex structure description
of it very manifest, and makes the Kahler structure description
implicit.

Note that for a generic quartic polynomial the
$K3$ we obtain is non-singular.  This is in sharp
contrast with the orbifold construction which led us
to 16 singular points.  It is possible to choose parameters
of deformation which lead to singular points for $K3$.
For example if we consider
$$z_1^4+z_2^4+z_3^4+z_4^4+4 z_1z_2z_3z_4=0$$
it is easy to see that the resulting $K3$ will have an $A_1$
singularity (one simply looks for non-trivial solutions to $dW=0$).

There are other ways to construct Calabi-Yau manifolds and in particular
$K3$'s.  One natural generalization to the above
construction is to consider weighted projective spaces
where the $z_i$ are identified under different rescalings.
In this case one considers quasi-homogeneous polynomials
to construct submanifolds.

\subsec{$K3$ as an elliptic manifold}
A certain subset of $K3$ manfiolds admit elliptic fibration.
What this means is that locally they look like a two torus
times a complex plane which we denote by $z$.
The complex structure of the torus varies
holomorphically as a function of $z$, and over $24$ points
on the $z$-sphere the elliptic fiber degenerates.
Let us describe this in more detail.

A 2 real dimensional torus (
also known as elliptic curve)
can be viewed as a double cover of the sphere
branched over 4 points.  Let us denote
the complex coordinate parametrizing the sphere by $x$.
Then an equation of the form
$$y^2=P_4(x)$$
defines a torus where
$P_4(x)$ is a polynomial of degree $4$ in $x$.  The fact
that this is a double cover of the plane is because for each
$x$ there are two values of $y$, related by a change in sign,
 satisfying the above equation.  The four branch points
are where $P_4(x)=0$, in which case we get only one
value of $y$ over it.  It is convenient to take one of the branch
points to infinity and define the torus through
$$y^2=x^3+f x +g$$
where with no loss of generality we have taken the coefficient
of $x$ and $y$ to be one, and we have eliminated the terms proportional
to $x^2$ by a shift in $x$.  The complex coefficients $f,g$ vary
the complex structure of the torus.  However the complex
structure of the two torus
is one dimensional.  This can be seen by the parallelogram
construction of the torus which up to rescaling we can choose
one length of the parallelogram to correspond to the point 1 on the
complex plane and the other to a complex parameter $\tau$.
If we rescale $(f,g)\rightarrow (\lambda^4 f,\lambda ^6 g)$,
the above equation is invariant if we redefine
$(x,y)\rightarrow (\lambda^2 x,\lambda^3 y)$.

Note that when the cubic $x^3+fx+g$ has double roots
the torus is degenerate.  This is because near such points
we can write the torus as $y^2=(x-x_0)^2$, which is a
one-dimension lower singularity of the $A_1$ type discussed
above.  This corresponds to a vanishing circle (instead
of the 2-sphere).  The discriminant of the cubic, where
the elliptic curve is singular is given by
$$\Delta =4f^3+27 g^2 =0$$

Now we come to the elliptic description of $K3$.
In view of the generality of the `fibration' picture,
let us be a bit more general.  Suppose we have a string
compactification $X$ whose data is parameterized by
a set of parameters ${\cal M}$ (i.e., its moduli space).
By fibering this over another space, we mean that these
parameters vary ``slowly'' over another space, known as
the base of the fibration.  In other words we let the parameters
of compactification on $X$, such as complex structure and radii,
become functions over the base.  Having said that, if we wish
to have an elliptic fibration over the sphere, all we have
to do is to make the $f$ and $g$ given in defining an
elliptic curve be functions of the sphere, which we denote
by the $z$ parameter.  In particular if we take $f(z)$ to be
a polynomial of degree $8$ and $g(z)$ to be a polynomial
of degree $12$ we obtain a $K3$ surface:
$$y^2=x^3+f(z) x +g(z)$$
 The condition that the degrees of $f$ and $g$ should
be in the ratio of $2$ to $3$ follows from the fact
that as $z\rightarrow \infty$ we generically wish to have
a non-singular elliptic curve (in which case for large $z$ we
can redefine $x$ and $y$ to see that we obtain a fixed
elliptic fibration at infinity).  Note the the discriminant
$\Delta$ is in this case a polynomial of degree 24.  This
means that over 24 points on the sphere the elliptic fiber
develops a degeneration of the form
$$xy=(z-z_0)$$
where the elliptic fiber is degenerate at $z=z_0$, where a
1-cycle (i.e. a circle) has shrunk to a point.  Note that the total
space including $(x,y,z)$ subject to the above equation is
not singular. In fact we can use $x$ and $y$ to coordinatize
this region and read off what $z$ is from the above equation.
It is only the fibration that becomes singular.  The situation
would be different, however if more than one fiber
becomes singular at the same point.  In particular if we have
$n$ singular fibers at the same point, i.e., if we locally
have
$$xy=(z-z_0)^n$$
we have an $A_{n-1}$ singularity of $K3$.

Note that the complex dimension of moduli of elliptic
$K3$'s is given by the number of parameters which
go into defining $f$ (9) plus the number of parameters
for $g$ (13) minus the $SL(2,C)$ action on the $z$ sphere
(3) minus the overall rescaling of the equation
defining elliptic curve (1), which gives us altogether
18 complex parameters.  The dimension of Kahler deformation
of $K3$ consistent with preservation of elliptic fibration is
2, one from the Kahler class (size) of the z-sphere, and one
from the Kahler class of the elliptic fiber.

\subsec{Moduli space of $K3$}
As discussed before $K3$ has $20$ complex deformations and
$20$ Kahler deformations.  To be more precise, if we consider
the moduli of metrics on $K3$ with $SU(2)$ holonomy this
space is 58 real dimensional space.  The way this arises is that we have
40 real parameters from the choice of complex structure and 20 real
parameters from the choice of Kahler structures.  However there
is a sphere worth of redundancy in that for every Ricci-flat
metric on $K3$ we can choose a sphere worth of choice of complex
structures consistent with it. Thus the space is 58 dimensional.
It turns out that this moduli space of geometric $K3$'s is
isomorphic to
$${\cal M}^{\rm geometric}=
{SO(19,3)\over SO(19)\times SO(3)\times SO(19,3;{\bf Z}) }\times R^{+}$$
The easiest way to understand these results is that if we consider
the elements of $H_2(K3)$, as discussed above, this space has
dimension 22, corresponding to the fact that $h^{2,0}=h^{0,2}=1$
and $h^{1,1}=20$.  Moreover since the dimensions of these cycles
is half the dimension of the total space we can consider an
intersection form on this 22 dimensional space corresponding
to the intersection between these cycles.  It turns out
this space is the same as the Narain lattice ${\Gamma}^{19,3}$.
For a fixed metric on $K3$ we can use the duality by the $*$-operation.
This gives a polarization on this lattice, exactly as the choice
of left- and right-moving momenta arise for toroidal compactifications.
It turns out that except for the overall rescaling of the $K3$ metric
which does not affect this polarization, the rest of the choice
of metric on $K3$ is faithfully represented by the change in the polarization
of the $H_2$ lattice and this gives rise to the above moduli space.  The
$R^{+}$ above corresponds to overall rescaling of the metric.

This is not the end of the story if we are considering string propagation
on $K3$.  In this case for each element of $H_2(K3)$ we can turn
on the fields $B_{\mu\nu}$. This space for $K3$ is 22 dimensional.
Thus we have 22 additional real parameters and that makes the total
dimension of moduli space $58+22=80$.  It turns out that this moduli
space is isomorphic to
$${\cal M}^{\rm stringy}={SO(20,4)\over SO(20)\times SO(4)\times SO(20,4;{\bf
Z})}$$ %
Note that the full homology lattice of $K3$ with the natural
intersection pairing is isomorphic to ${\Gamma}^{20,4}$ lattice.
This result on stringy moduli space of
$K3$ can be interpreted by saying that for strings
the polarization on the full homology (not just the middle homologies)
are relevant.

For elliptic $K3$'s as noted above the moduli space is 18 complex dimensional
plus 2 real parameters corresponding to the Kahler classes of the base and
fiber.  This space is isomorphic to
$${\cal M}^{\rm elliptic}={SO(18,2)\over SO(18)\times SO(2)\times
SO(18,2;{\bf Z})}\times R^+\times R^+$$
This result can be interpreted by noting that the choice of the elliptic
fiber amounts to choosing a null vector in the $H_2$ lattice, which
has to be preserved by the elliptic metrics on $K3$ and this reduces
the group acting on moduli space
from $SO(19,3)$ to $SO(18,2)$ via a null
reduction (note that the self-intersection of the elliptic fiber is zero
and thus corresponds to a null vector).

\subsec{$N=2$ theories on $K3$}
Let us now consider $N=2$ theories compactified on $K3$.
The two choices will lead to inequivalent theories in 6 dimensions,
this is because the chirality of the covariantly constant spinor
is correlated with the 6 dimensional chirality.  Thus if we start
with the chiral $N=2$ theory as is the case for type IIB strings,
we end up with a chiral theory in 6 dimensions, whereas if we
start with the non-chiral type IIA theory we will end up with the
non-chiral theory in 6 dimensions, which has exactly the same supersymmetry
as the $N=1$ theories on $T^4$.

Let us first consider the moduli space of type IIA on $K3$.  Recall
that the bosonic fields are $g_{\mu \nu},B_{\mu\nu},\phi,A_\mu,C_{\mu
\nu \rho}$.  The choices of $g_{\mu\nu}$ and $B_{\mu \nu}$ on $K3$
lead to the stringy moduli of $K3 $ discussed above.  Since $H_1(K3)$
and $H_3(K3)$ are trivial we cannot make any choice for the vacuum solutions
for $A_\mu$ or $C_{\mu\nu\rho}$ fields (which preserve supersymmetry
with zero field strength).  We still have the choice of string coupling
which is a positive real parameter. Thus the moduli space is
$${\cal M}^{IIA}={SO(20,4)\over SO(20)\times SO(4)\times SO(20,4;{\bf Z})}
\times R^+$$
Note that this is {\it exactly} the same as the moduli space
of $N=1$ theories (type I or heterotic theory) on $T^4$.
As we will discuss later it
turns out that this is not accidental!

As for type IIB on $K3$ in addition to the stringy moduli space
of $K3$ and the dilaton
we note that we have the choice for the $\chi
,{\tilde B}_{\mu\nu} $ and $D_{\mu\nu\rho\lambda}$ coming
from the RR sector.  The choices for the vevs of these fields
with zero field strength correspond to even cohomology elements
of $K3$ and so there are 24 parameters.  Thus the total
moduli space has dimension $80+1+24=105$.  It turns out
that by supersymmetry arguments this full moduli space is simply
given by
$${\cal M}^{\rm type IIB}={SO(21,5)\over SO(21)\times SO(5)\times
SO(21,5;{\bf Z})}$$
(where the integral modding out in the denominator is conjectural).

\subsec{N=1 theories on $K3$}
We now consider compactification of $N=1$ theories on
$K3$.  Recall that the bosonic fields for them is given
by $g_{\mu\nu},B_{\mu\nu},\phi,A_\mu$ where the gauge
field belongs to $SO(32)$ group or $E_8\times E_8$.
As far as the moduli for $g_{\mu \nu}$ and $B_{\mu \nu}$ are
concerned we get the same story as in the type II compactifications.
Now, however we can choose non-vanishing configuration
for the gauge fields without breaking supersymmetry.  The reason
is that if we consider gauge field configurations on $K3$ satisfying
self-duality configuration:
$$F=*F$$
then they preserve the same supersymmetry as the metric
configuration on $K3$.  These configurations of gauge fields
are of course the standard instantons of gauge theory.  In fact
not only we {\it can} turn on these gauge fields we {\it have to}
turn them on if we wish to have a consistent solution.  This is
because the $N=1$ theories have an equation of motion of the form
\eqn\inst{dH =Tr F\wedge F- Tr R\wedge R}
where $H=dB$ is the field strength for the anti-symmetric field.
This equation implies that, in the absence of singularities for the
$H$ field, since the integral over $K3$ of the left hand side vanishes,
the instanton number of the gauge field should be related
to a topological computation involving the curvature, which
turns out to be $-{1\over 2}p_1(R)$ where $p_1$ is the first
Pontrjagin class.  For $K3$ this is 24.  We thus learn
that we have to consider instanton number 24 configuration
for the gauge field.  For $E_8\times E_8$ this
total instanton number can be distributed among the
two $E_8$'s, with instanton numbers $(12+n,12-n)$.

In principle we can also consider configurations
where the integral of $dH$ over $K3$ is not zero,
i.e.,
$$\int_{K3} dH =m$$
The meaning of this singularity will be discussed
in the next section where we will discuss p-branes.
Let us just state that what this configuration means
is that we have $m$ 5-branes.  If we choose this, then
we have $24-m$ instantons to put on the gauge groups.

The moduli space for compactification of $N=1$ theories
will in addition to the moduli describing the geometry
of $K3$ includes moduli describing the choices
of gauge bundle.  This makes the description more complicated
and we will not discuss it here.

\subsec{Calabi-Yau threefolds}
We have already noted that the next case of interest
involves compactification on manifolds of $SU(3)$ holonomy.
This preserves 1/4 of the supersymmetry.  In particular if
we compactify $N=2$  theories on Calabi-Yau threefolds we obtain
$N=2$ theories in $d=4$, whereas if we consider $N=1$  theories
we obtain $N=1$ theories in $d=4$.  In the latter case, as
discussed in the context of $K3$ compactifications we have
to choose the instanton class of the bundle to be in the same
class as the tangent bundle on the manifold.
 Let us first consider some
simple classes of Calabi-Yau threefolds and then we discuss
some aspects of them in the cases of $N=2$ and $N=1$ theories.

Calabi-Yau threefolds are manifolds with $SU(3)$ holonomy.
In particular if we wish to construct them as toroidal
orbifolds we need to consider six dimensional tori, three complex
dimensional,  which
have discrete isometries residing in $SU(3)$ subgroup of the $O(6)=SU(4)$
holonomy group.  A simple example is if we consider the product
of three copies of $T^2$ corresponding to the Hexagonal lattice
and mod out by a simultaneous ${\bf Z}_3$ rotation on each
torus (this is known as the `Z-orbifold'). This $Z_3$ transformation
has $27$ fixed points which can be blown up to give rise to a smooth
Calabi-Yau.
 We can also
consider description of Calabi-Yau threefolds in algebraic
geometry terms for which the complex deformations of the manifold
can be typically realized as changes of coefficients of defining
equations, as in the $K3$ case.  For instance we can consider
the projective 4-space ${\bf P}^4$ defined by 5 complex not
all vanishing coordinates $z_i$ up to overall rescaling,
and consider the vanishing locus of a homogeneous degree 5 polynomial
$$P_5(z_1,...,z_5)=0$$
This defines a Calabi-Yau threefold, known as the quintic three-fold.
This can be generalized to the case of product of several projective
spaces with more equations. Or it can be generalized
by taking the coordinates to have different homogeneity weights.
This will give a huge number of Calabi-Yau manifolds.

We can also consider Calabi-Yau manifolds which admit elliptic
fibrations, as was the case for elliptic $K3$'s.  For example
we can consider a space which as the base  contains ${\bf P}^1
\times {\bf P}^1$, with an elliptic fiber over it.  Let
us denote the coordinates of ${\bf P}^1\times {\bf P}^1$ by
$(z_1,z_2)$.  Then we can write the equation denoting
the complex structure of the elliptic fiber by
$$y^2=x^3+f_{(8,8)}(z_1,z_2)x +g_{(12,12)}(z_1,z_2).$$
where $f_{(8,8)}$ and $g_{(12,12)}$ are polynomials of
bidegree 8 and 12 in $(z_1,z_2)$ respectively.

In various applications it is important to study
the moduli space of Calabi-Yau manifolds.  This generically
splits to complex deformations and Kahler deformations.
The number of Kahler deformations is given by $h^{1,1}$ of
the manifold corresponding to the choice of the Kahler form.
Together with the choice of the $B_{\mu\nu}$ field the dimension
of the Kahler deformation is given by $h^{1,1}$ complex parameters.
The number of complex deformations is given by $h^{2,1}$ of
the Calabi-Yau and is typically realized by complex
coefficients in the defining equations.  Let us enumerate
these for the examples of Calabi-Yau manifolds presented above.

For the Z-orbifold, from the untwisted sector, we obtain
$h^{1,1}_u=9$ corresponding to the choice $dz_i\wedge d{\overline z}_j$
for $i,j=1,2,3$, and from the twisted sector we obtain $h^{1,1}_t=27$
additional contributions (corresponding to blowing up the singularity
at each of the 27 fixed points, which locally replaces the neighborhood
of the fixed point with a line bundle over ${\bf P}^2$ which has
$h^{1,1}=1$), we thus obtain altogether $h^{1,1}=36$ parameters.
For Z-orbifold $h^{2,1}=0$ and thus there are no complex
deformations.  This is easy to see because the $Z_3$ symmetry
exists only for a unique complex structure on $T^2\times T^2\times T^2$
(and there are no additional complex structure parameters
arising from the twisted sector).

For the example of quintic threefold it is easy to count the complex
deformations realized as coefficients of a degree 5 homogeneous
polynomial.  This gives 101 complex parameters, which is in agreement
with $h^{2,1}=101$.  The dimension of $h^{1,1}=1$ which means
that there is only one Kahler class, and that corresponds to the
overall size of the quintic.
Finally for the elliptically fibered Calabi-Yau, over
${\bf P}^1\times {\bf P}^1$, we have $h^{1,1}=3$ coming
from the three Kahler classes of the size of each ${\bf P}^1$
plus the size of the elliptic fiber.  To count $h^{2,1}$ note that
there are $9\cdot 9$ coefficients in $f_{(8,8)}$ and
$13\cdot 13$ coefficients in defining $g_{(12,12)}$, and subtracting
the $SL(2,{\bf C})$ symmetry of each ${\bf P}^1$ and the overall
rescaling of the equation, we find $9\cdot 9+13\cdot 13-3\cdot 2-1=243$.
This in particular means that $h^{2,1}=243$.

\subsec{N=2 theories on Calabi-Yau 3-folds}
As mentioned above if we consider N=2 theories on Calabi-Yau
threefolds we obtain $N=2$ supersymmetric theories in four dimensions.
An $N=2$ quantum field theory contains hypermultiplets and vector
multiplets in addition to the $N=2$ supergravity multiplet.
If we consider type IIA strings on Calabi-Yau threefolds
the number of vector multiplets is $h^{1,1}$ and the number
of hypermultiplets is $h^{2,1}$.  To see this note that
expectation values for the scalar in the vector
multiplet contain a complex field which can be identified
with the choice of the Kahler class and the choice of the
NS-NS $B_{\mu\nu}$ in the internal space. Similarly
the four component expectation values for a hypermultiplet
can be identified with the complex structure of the manifold
($h^{2,1}$ complex variables) together with the expectation
values of the anti-symmetric three form ($h^{2,1}$).  Two
more choices for the antisymmetric three form ($h^{3,0}+h^{0,3}$)
as well as the choice of the string coupling and the dual
to the spacetime components of $B_{mu\nu}$ (the axion) forms
an extra hypermultiplet, we thus end up with $h^{1,1}$ vector
multiplets and $h^{2,1}+1$ hypermultiplets.  If we consider
instead type IIB strings the role of complex and Kahler
structures of Calabi-Yau are exchanged and we end up with
$h^{2,1}$ vector multiplets and $h^{1,1}+1$ hypermultiplets.

An important aspect of type IIA and type IIB strings on Calabi-Yau
threefolds is the notion of mirror symmetry.  The simplest
way to view this is to note that many Calabi-Yau threefolds
can be viewed (roughly speaking) as a fibered space, with
a $T^3$ fiber and an $S^3$ base.  The inversion of the volume
of the $T^3$ fiber leads to another Calabi-Yau manifold which we
identify with the mirror Calabi-Yau.  In this way the
complex and Kahler structures get exchanged (note that
this is the natural generalization of the 1-fold case
where we consider an $T^1=S^1$ fiber over $S^1$ and do scale
inversion on the fiber $T^1$).  If we exchange a Calabi-Yau
manifold with its mirror we have the exchange of
$$h^{1,1}\leftrightarrow h^{2,1}$$
In particular type IIA on one Calabi-Yau is equivalent
to type IIB on its mirror.

\subsec{N=1 strings on Calabi-Yau threefolds}
If we consider $N=1$ strings (i.e. heterotic strings or Type I
strings) on Calabi-Yau threefolds we obtain an $N=1$ supersymmetric
theory in four dimensions.  As discussed in the context of $K3$
compactification we have to choose a guage field configuration
whose instanton class ${\rm Tr}F\wedge F$ is in the same
class as (minus one half) the pontrjagin class of the Calabi-Yau threefold
which is proportional to ${\rm Tr} R\wedge R$.
There are many ways to fulfill this.  One particularly
simple way is to consider an $SU(3)$ subgroup of the gauge
group and choose the gauge connection to be the same
as the spin connection on the threefold.  For example,
in the context of $E_8\times E_8$ heterotic strings, if
we choose an $SU(3)$ subgroup of one of the $E_8$'s
we will end up breaking it to $E_6$.  Moreoever the charged matter
are in the $27$ and $\overline{27}$ representations of $E_6$.
The number of $27$'s is given by $h^{1,1}$ and those of $\overline{27}$
by $h^{1,2}$.  There will also be many neutral fields. In particular
 $h^{1,1}+h^{2,1}$ of them will correspond
to deforming the (complexified) Kahler structure and
the complex structure of the threefold. There will also
be some neutral scalar fields which correspond to deforming
the gauge connection away from its identity with spin connection
(note that any continuous deformation of the gauge connection
will still be consistent with the topological condition
that instanton class of the bundle agree with half the pontrjagin
class of the manifold).  In the $N=1$ case, as is well known
there could be non-perturbative superpotentials generated
and the above spectrum is the spectrum of massless modes
{\it before} taking into account such corrections, which
could in principle give mass to some of them.

\newsec{Solitons and String Theory}
Solitons arise in field theories when the vacuum configuration
of the field has a non-trivial topology which allows non-trivial
wrapping of the field configuration at spatial infinity around
the vacuum manifold.  These will carry certain topological charge related
to the `winding' of the field configuration around the vacuum configuration.
Examples of solitons include magnetic monopoles in four
dimensional non-abelian
gauge theories with unbroken $U(1)$, cosmic strings and domain
walls.
The solitons naively play a less fundamental role than
the fundamental fields which describe the quantum field theory.
In some sense we can think of the solitons as `composites' of
more fundamental elementary excitations.  However as is well
known, at least
in certain cases,
this is just an illusion.
In certain cases it turns out that we can reverse the
role of what is fundamental and what is composite by
considering a different regime of parameter.  In such
regimes the soliton may be viewed as the elementary
excitation and the previously viewed elementary
excitation can be viewed as a soliton.
A well known example of this phenomenon happens
in 2 dimensional field theories.
Most notably the boson/fermion equivalence in the two dimensional
sine-Gordon model, where the fermions may be viewed as solitons
of the sine-Gordon model and the boson can be viewed as a composite
of fermion-anti-fermion excitation.  Another example is the
T-duality we have already discussed in the context of 2d worldsheet
of strings which exchanges the radius of the target space
with its inverse.  In this case the winding modes may
be viewed as the solitons of the more elementary excitations
corresponding to the momentum modes.  As discussed before
$R\rightarrow 1/R$ exchanges momentum and winding modes.
In anticipation of generalization of such dualities to
string theory, it is thus important to study various
types of solitons that may appear in string theory.

As already mentioned solitons typically carry some conserved
topological charge.  However
in string theory every conserved charge
is a gauge symmetry.  In fact this is to be expected
from a theory which includes quantum gravity.  This
is because the global charges of a black hole will have
no influence on the outside and by the time the black hole
disappears  due to Hawking radiation, so does the global charges it may carry.
So the
process of formation and evaporation of black hole leads to
a non-conservation of global charges.  Thus for any soliton,
its conserved topological charge must be a gauge charge.  This
may appear to be somewhat puzzling in view of the fact
that solitons may be point-like as well as string-like, sheet-like
etc.  We can understand how to put a charge on a point-like
object and gauge it.  But how about the higher dimensional
extended solitonic states? (note that if we
view the higher dimensional solitons as made of point-like structures
the soliton has no stability criterion as the charge can
disintegrate into little bits)

Let us review how it works for point particles (or point solitons):
We have a 1-form gauge potential $A_{\mu}$ and the coupling
of the particle to the gauge potential involves weighing
the worldline propagating in the spacetime with background
$A_{\mu}$ by
$$Z\rightarrow Z {\rm exp}(i \int_\gamma A)$$
where $\gamma$ is the world line of the particle.  The gauge principle
follows from defining an action in terms of $F=dA$:
\eqn\acti{S=\int F\wedge *F}
where $*F$ is the dual of the $F$, where we note that shifting
$A\rightarrow d\epsilon$ for arbitrary function $\epsilon$ will
not modify the action.
Suppose we now consider instead of a point particle a
$p$-dimensional extended object.  In this convention $p=0$
corresponds to the case of point particles and
$p=1$ corresponds to strings and $p=2$
corresponds to membranes, etc.  We shall
refer to $p$-dimensional extended objects as
$p$-branes (generalizing `membrane').  Note that the worldvolume
of a $p$-brane is a $p+1$ dimensional subspace $\gamma_{p+1}$
of spacetime.  To generalize what we did for the case of point particles
we introduce a gauge potential which is a $p+1$ form $A_{p+1}$
and couple it to the charged $p+1$ dimensional state by
$$Z\rightarrow Z {\rm exp}(i \int_{\gamma _{p+1}} A_{p+1})$$
Just as for the case of the point particles we introduce
the field strength $F=dA$ which is now a totally antisymmetric
$p+2$ tensor.  Moreover we define the action as in \acti ,
which possesses the gauge symmetry $A\rightarrow d \epsilon$
where $\epsilon$ is a totally antisymmetric tensor of rank $p$.

\subsec{Magnetically Charged States}
The above charge defines the generalization
of electrical charges for extended objects.  Can
we generalize the notion of magnetic charge?
Suppose we have an electrically charged particle
in a theory with spacetime dimension $d$.
Then we measure the electrical charge by surrounding
the point by an $S^{d-2}$ sphere and integrating $*F$ (which
is a $d-2$ form) on it, i.e.
$$Q_E=\int_{S^{d-2}} *F$$
Similarly it is natural to define the magnetic charge.  In the
case of $d=4$, i.e. four dimensional spacetime, the magnetically
charged point particle can be surrounded also by a sphere and the
magnetic charge is simply given by
$$Q_M=\int_{S^2} F.$$
Now let us generalize the notion of magnetic charged states
for arbitrary dimensions $d$ of spacetime and arbitrary
electrically charged $p$-branes.  From the above description
it is clear that the role that $*F$ plays in measuring
the electric charge is played by $F$ in measuring the magnetic
charge.  Note that for a $p$-brane $F$ is $p+2$ dimensional,
and $*F$ is $d-p-2$ dimensional.  Moreover note that a sphere
surrounding a $p$-brane is a sphere of dimension $d-p-2$.  Note
also that
for $p=0$ this is the usual situation.  For higher $p$, a
$p$-dimensional subspace of the spacetime is occupied by the
extended object and so the position of the object is denoted
by a point in the transverse $(d-1)-p$ dimensional space
which is surrounded by an $S^{d-p-2}$ dimensional sphere.

Now for the magnetic states the role of $F$ and $*F$ are exchanged:
$$F\leftrightarrow *F$$
To be perfectly democratic we can also define a magnetic
gauge potential $\tilde A$ with the property that
$$d\tilde A=*F=*dA$$
In particular noting that $F$ is a $p+2$ form
 we learn that $*F$ is an $d-p-2$ form and thus $\tilde A$
is an $d-p-3$ form.  We thus
deduce that the magnetic state will be an $d-p-4$-brane (i.e.
one dimension lower than the degree of the magnetic
gauge potential ${\tilde A}$).
Note that this means that if we have an electrically charged
$p$-brane, with a magnetically charged dual $q$-brane then
we have
\eqn\mage{p+q=d-4}
This is an easy sum rule to remember.
Note in particular that for a 4-dimensional spacetime an electric
 point charge ($p=0$) will have a dual magnetic point charge ($q=0$).
Moreover this is the only spacetime dimension where both the electric
and magnetic dual can be point-like.

Note that a p-brane wrapped around an r-dimensional compact
object will appear as a $p-r$-brane for the non-compact
spacetime.  This is in accord with the fact that if we decompose
the $p+1$ gauge potential into an $(p+1-r)+r$ form consisting
of an $r$-form in the compact direction we will end up with an
$p+1-r$ form in the
non-compact directions.  Thus the resulting state is charged
under the left-over part of the gauge potential.  A
particular case of this is when $r=p$ in which case
we are wrapping a p-dimensional extended object about
a p-dimensional closed cycle in the compact directions.
This will leave us with point particles in the non-compact
directions carrying ordinary electric
charge under the reduced gauge potential which now is a 1-form.

\subsec{String Solitons}
{}From the above discussion it follows that the charged states
will in principle exist if there are suitable gauge potentials
given by $p+1$-forms.  Let us first consider type II strings.
Recall that from the NS-NS sector we obtained an anti-symmetric
2-form $B_{\mu\nu}$.  This suggests that there is a 1-dimensional
extended object which couples to it by
$${\rm exp}(i\int B)$$
But that is precisely how $B$ couples to the worldsheet of the
fundamental string.  We thus conclude that {\it the fundamental
string carries electric charge under the antisymmetric
field} $B$.  What about the the magnetic dual to the fundamental
string?  According to \mage\ and setting $d=10$ and $p=1$ we
learn that the dual magnetic state will be a 5-brane.  Note that
as in the field theories we expect that in the perturbative
regime for the fundamental fields, the solitons be very massive.
This is indeed the case and the 5-brane magnetic dual can
be constructed as a solitonic state of type II strings with
a mass per unit 5-volume going as $1/g^2$ where $g$ is the
string coupling.   We shall discuss below some relationship
between these 5-branes and the type II theories on ALE space
with $A_n$ singularities.

Let us also recall from our discussion in section 1
that type II strings also have anti-symmetric fields coming
from the R-R sector.  In particular for type IIA strings
we have 1-form $A_{\mu}$ and 3-form $C_{\mu\nu \rho}$ gauge
potentials. Note that the corresponding magnetic
dual gauge fields will be 7-forms and 5-forms respectively
(which are not independent degrees of freedom).  We can
also include a 9-form potential which will have trivial
dynamics in 10  dimensions.  Thus it is natural
to define a generalized gauge field ${\cal A}$ by taking the sum over
all odd forms and consider the equation ${\cal F}=*{\cal F}$
where ${\cal F}=d{\cal A}$.  A similar statement
applies to the type IIB strings where from the R-R
sector we obtain all the even-degree gauge potentials
(the case with degree zero can couple to a ``$-1$-brane'' which
can be identified with an instanton, i.e. a point in spacetime).
We are thus led to look for p-branes with even $p$ for
type IIA and odd $p$ for type IIB which carry charge under
the corresponding RR gauge field.  It turns out
that surprisingly enough the states in the elementary
excitations of
string all are neutral under the RR fields.  We are thus
led to look for solitonic states which carry RR charge.
Indeed there are such p-branes and they
are known as $D$-branes, as we will now review.

\subsec{D-branes: The carriers of RR charge}
In the context of field theories constructing
solitons is equivalent to solving classical field equations
with appropriate boundary conditions.  For string theory
the condition that we have a classical solution is equivalent
to the statement that propagation of strings in
the corresponding background
would still lead to a conformal theory on the worldsheet
of strings, as is the case for free theories.

In search of such stringy p-branes, we are thus
led to consider how could a p-brane modify the
string propagation.  Consider an $p+1$ dimensional
plane, to be identified with the worldvolume of the $p$-brane.
Consider string propagating in this background.  How
could we modify the rules of closed string propagation
given this $p+1$ dimensional sheet?  The simplest
way turns out to allow closed strings to open up
and end on the $p+1$ dimensional worldvolume.  In other
words we allow to have a new sector in the theory
corresponding to open string with ends lying on this
$p+1$ dimensional subspace.  This will put Dirichlet
boundary conditions on $10-p-1$ coordinates of string
endpoints.  Such $p$-branes are called $D$-branes,
with D reminding us of Dirichlet boundary conditions.
In the context of type IIA,B we also have to specify
what boundary conditions are satisfied by fermions.
This turns out to lead to consistent boundary conditions
only for $p$ even for type IIA string and $p$ odd for
type IIB.\foot{Note that this is a consequence of the fact
that for type IIA(B), left-right exchange is a symmetry
only when accompanied by a $Z_2$ spatial reflection with
determinant -1(+1).}
 Moreover it turns out that they do carry
the corresponding RR charge.

Quantizing the new sector of type II strings
in the presence of D-branes is rather straight
forward. We simply consider the set of oscillators
as before, but now remember that due to the Dirichlet
boundary conditions on some of the components of string
coordinates, the momentum of the open string lies on
the $p+1$ dimensional worldvolume of the D-brane.
It is thus straight forward to deduce that
the massless excitations propagating on the D-brane
 will lead to the dimensional reduction
of $N=1$ $U(1)$ Yang-Mills from $d=10$ to $p+1$ dimensions.
In particular the $10-(p+1)$ scalar fields living
on the D-brane, signify the D-brane excitations
in the $10-(p+1)$ transverse dimensions.
This tells us that the significance of the new open string
subsector is to quantize the D-brane excitations.

An important property of D-branes is that when
$N$ of them coincide we get a $U(N)$ gauge theory
on their worldvolume.  This follows because we have
$N^2$ open string subsectors going from one D-brane to another
and in the limit they are on top of each other all
will have massless modes and we thus obtain the reduction
of $N=1$ $U(N)$ Yang-Mills from $d=10$ to $d=p+1$.

If we consider the tension of D-branes they
are proportional to $1/g$ where $g$ is the string coupling
constant.  Note that as expected at weak coupling
they have a huge tension.  An important property
of D-branes is that they are BPS states.  A BPS state
is a state which preserves a certain number of supersymmetries
and as a consequence of which one can show that their
mass (per unit volume) and charge are equal.  This in particular
guarantees their absolute stability against decay.

\subsec{Fivebranes and ALE spaces}
In the context of $K3$ compactification we discussed
$A_{n-1}$ singularities as corresponding to $n$ cosmic strings
coming together on a point $z_0$ on the base
 $z$-sphere.  Let us consider a three cycle $C$ consisting of a circle
$\gamma$
around $z_0$ and the $T^2$ fiber above it.  Note that as we go
around the cycle the complex modulus of $T^2$ shifts by $\tau \rightarrow
\tau +n$.  Now let us do T-duality on the cycle of $T^2$ which
vanishes as we approach $z_0$.  In this way we exchange
the role of complex moduli of torus with its
(complexified) Kahler modulus.  In particular as we go around $\gamma$
now we have $B\rightarrow B+n$ where $B$ corresponds
to the antisymmetric field with components along $T^2$.  This implies
that
$$\int_C dB=n$$
which shows that we now have a source of $n$-units of {\it magnetic}
charge for $B$, i.e. n units of 5-brane charge.  Note however
in showing this relation we used $R\rightarrow 1/R$ once, and
so this exchanges type IIA with type IIB.  We thus learn
that $n$ 5-branes of type IIA(B) is equivalent to type IIB(A)
in the presence of an $A_{n-1}$ ALE singularity.

{}From these facts we can also deduce some aspects of the theory
living on the fivebrane
for type IIA and type IIB strings.  In particular we learn
that for type IIB 5-brane, the bosonic fields include a $U(1)$ vector
field together with 4 scalars, and for type IIA 5-brane, the bosonic
fields include an antisymmetric two form with self-dual field strength,
and five scalars.

\subsec{$N=1$ Theories and p-branes}
For $N=1$ theories, the only antisymmetric gauge potential is
the two-form $B_{\mu\nu}$.  We thus expect a 1-brane and a
dual 5-brane.
For heterotic string theory the 1-brane is identified with the
closed string itself.  In the type I case the 1-brane comes
from the RR sector and is not identified with the fundamental string
(note that the closed string for type I can decay to open string and
thus does not couple to a conserved charge).  The carrier of the 1-brane
charge is a D1-brane.  Moreover the dual magnetic state is a D5
brane.  From the viewpoint of D-branes, the existence of the
open string subsector for type I strings with
32 Chan-Paton indices can be viewed as follows:  We consider
the orientifold of type IIB.  This induces a tadpole for
a 10-form gauge potential (which is not dynamical) from the RR-sector
which needs to be canceled by placing 32 9-branes. The
strings ending on the 32 D9-branes is the open string sector
of type I strings.

\subsec{D-branes and T-duality}
Consider type II strings with a
D-brane wrapped around a circle
and consider doing T-duality along the circle. We are interested
in knowing what happens to D-brane after doing T-duality.
 Note
that this exchanges type IIA and B, and thus the dimension
of D-brane should change by an odd number.  In fact what happens
is that it goes down by one.  This is because under T-duality
the Dirichlet and Neumann boundary conditions get exchanged.
Thus if a D-brane is wrapped around the circle the open strings
ending on it have
Neumann boundary conditions along the circle and after T-duality
they become Dirichlet boundary condition, thus decreasing the
worldvolume dimension of the D-brane by one unit.  Similarly
if the D-brane was not wrapped around the circle, and its
position was represented by a point on the circle, after T-duality
the Dirichlet boundary conditions turn to Neumann and we end
up with one higher dimensional D-brane.

For type I string T-duality works in a slightly more delicate way:
For example  consider compactification of type I string on a circle.  We
take the radius of the circle to very small size, and by T-duality
we obtain a theory which now has 32 D8-branes, instead of 32 D9-branes
(as discussed above).  Their position is labeled by 32 points which
are symmetric under the $Z_2$ reflection on a circle and moving
the positions correspond
to changing the Wilson lines of the $SO(32)$ theory along the circle.
  But there are also two additional special points which are the points
along the circle fixed by the $Z_2$ involution.  The best way to understand
this theory is to consider type IIA theory compactified on a circle
and mod out by the orientifold symmetry which at the same time
acts by a reflection on the circle.  The two fixed points of this
action are known as the orientifold plane.  When the 32 D8 branes are
on one or the other orientifold plane we have the $SO(32)$ gauge symmetry
restored.  In the generic position
of the D8-branes that will be broken to $U(1)^{16}$.

\newsec{Applications of D-branes}

We have learned that type IIA,B strings admit a collection
of D-branes.  It is natural to ask what kinds of properties
they lead to.

We have discussed that in compactification of string
theory we often end up with singular limits of manifolds
when some cycles shrink to zero size. A typical example
is when an $n$-dimensional sphere shrinks to zero size.
What is the physical interpretation of this singularity?

Suppose we consider for concreteness an $n$-dimensional
sphere $S^n$ with volume $\epsilon\rightarrow 0$.
Then the string perturbation theory breaks down when
$\epsilon << \lambda$ where $\lambda$ is the string coupling
constant.  If we have  $n$-brane solitonic states
such as D-branes then
we can consider a particle solitonic state corresponding
to wrapping the $n$-brane on the vanishing $S^n$.  The mass of this state
is proportional to $\epsilon$, which implies that in the limit
$\epsilon \rightarrow 0$ we obtain a massless soliton.
An example of this is when we consider type IIA compactification
on $K3$ where we develop an $ADE$ singularity. Then by wrapping
D2-branes around vanishing $S^2$'s of the ADE singularity
we obtain massless states, which are vectors and charged
under the Cartan $U(1)^r$ of the ADE.  This in fact implies
that in this limit we obtain enhanced $ADE$ gauge symmetry.
Had we been considering type IIB on $K3$ near an $ADE$ singularity,
the lightest mode would be obtained by wrapping a D3-brane
around vanishing $S^2$'s, which leaves us with a string state with
tension of the order of $\epsilon$. The dynamics in this regime
is not well understood, as the light degrees of freedom
are string-like rather than particle-like.  This kind
of regime which exists in other examples of compactifications
as well is called the phase with {\it tensionless strings}\foot{We could
have considered higher dimensional D-branes wrapping around
the vanishing cycles, but in such cases by dimensional analysis
one can see that the relevant mass scale would be smallest for
the smallest dimension D-brane.}.

Another interesting example along the lines above occurs
when we consider Calabi-Yau threefold compactification of type IIB
strings, in the presence of vanishing $S^3$'s.  If an
$S^3$ vanishes we obtain a massless particle by wrapping
a D3 brane around $S^3$, which is a charged hypermultiplet
of an $N=2$ supersymmetric $U(1)$ theory.    Suppose we
have for instance 2 vanishing $S^3$'s as we change the
scalar vev for the $U(1)$ vector multiplet (which controls
the size of the $S^3$'s).  Then we obtain a $U(1)$ theory
with two charged massless hypermultiplets.  We can then
consider Higgsing the $U(1)$ and we are left with one
massless hypermultiplet.  This simple physical
process has an interesting mathematical application:
It corresponds to replacing the two shrunk $S^3$'s by
two growing $S^2$'s!  In other words the topology
of the Calabi-Yau has changed and we obtain a manifold
with one less $h^{2,1}$ and one more $h^{1,1}$.  In this
way a singular transition from one manifold to the other
can be given a smooth and simple physical description.

In the above applications the D-brane is mainly wrapped
about the internal direction and as far as the non-compact
spacetime is concerned it behaves as a point-like or sometimes
string-like object.  However there is another useful application
of p-branes and that is in mimicking the singularities of the
compactified spacetime.  In this case the p-brane will be
filling the spacetime.  One particularly simple application
of this is as follows:
By the discussion about the relation between
5-branes and $A_{n-1}$ singularities, since type IIA
in the presence of $A_{n-1}$ singularity acquires $SU(n)$
gauge symmetry we thus conclude, by T-duality,
that type IIB in the presence of $n$ coinciding 5-branes acquires
also $SU(n)$ gauge symmetry in the 5+1 dimensional
worldvolume of the coinciding 5-branes.  This sounds very much
like what we would have expected if instead of the 5-brane
associated to the $NS-NS$ $B$ field, we were considering
Dirichlet 5-brane.  This is the first hint that there is
a symmetry exchanging the two.

\subsec{Counting Degeneracy of D-branes: Applications
to Black-hole physics}

Let us consider type IIA theory compactified on $K3$.  Let us
consider how many D2-branes (Dirichlet 2-branes) are there
with some specific charges. In other words we are interested
in counting the number of BPS states with specific charges
given by the choice of an element in the second homology
of $K3$ which represents the homology class of a D2-brane.
Instead of being general let us consider an elliptic
$K3$ and consider D2-branes having specific
charges, corresponding to the wrapping of the D2-brane
$n$-times over the elliptic fiber and once over the base.

In fact it is quite simple to find the appropriate
configuration of such D2-brane:  It consists
of a degenerate Riemann surface consisting of
a sphere (the base of the elliptic fibration) together
with $n$ tori connected to it over $n$ points on the sphere.
However there is a moduli space of such configurations:  In
fact we can move the $n$ points as we wish over the base sphere.
Moreover on a D2-brane lives a $U(1)$ gauge field and so as far
as ground state configurations are concerned we can consider
wilson lines of the $U(1)$ over this Riemann surface,
which corresponds to choosing wilson lines on each of the
$n$-tori.  The choice of a $U(1)$ Wilson line on a torus
is equivalent to the choice of a point on the dual torus (which
can be easily understood by applying T-duality to the torus
turning the D2-brane to D0-brane). Thus as we move the base
points together with the wilson line on the $T^2$ above it, we
span the (dual) $K3$.
  Thus we find that the moduli
of configuration for the above D2-brane corresponds to the choice
of $n$ points on $K3$.  However the order of the points is irrelevent
and so we can write
$${\cal M}_{1,n}={\rm Sym}^n K3$$
where ${\cal M}_{1,n}$ denotes the moduli space of
the configurations of D2-brane with charges (1,n) corresponding
to wrapping about the base and fiber, and ${\rm Sym}^n K3$ denotes
the n-fold symmetric product of $K3$, i.e. $K3^{\otimes n}/S_n$ where $S_n$
is the permutation group on $n$-objects.
To find the number of $D2$ branes we have to quantize the above
moduli space.  The analogy to keep in mind is that the moduli
space of a point particle in a box is given by the choice of a point
in the box, and quantizing that means choosing momentum modes,
the ground states would correspond to zero momentum states.  In
the supersymmetric case that we are considering the number
of ground states of this quantum system is given by the cohomology
of ${\cal M}_{1,n}$.  It is relatively simple
to count the dimension of cohomology for ${\cal M}_{1,n}$
using orbifold realization of symmetric products
(which tells us how to deal with the singularities
of the space).  For $n=1$ we simply get the cohomology
of $K3$ which is $24$ dimensional.   Let us symbolically
denote these states by
$$|i\rangle={\alpha}^i_{-1}|0\rangle$$
where $i=1,...24$.
For $n=2$, we should consider the ground states of $K3\times K3/Z_2$.
{}From the untwisted sector we obtain the symmetrization
of two copies of the above state, i.e.
$$|(i,j)\rangle={\alpha}^i_{-1}\alpha^j_{-1}|0\rangle$$
where writing it in terms of ``creation operators''
has the advantage of making it manifest that we symmetrize
over them.  However now we also have the twisted sector
where the two $K3$'s coincide.  The ground states of the twisted
sector consists thus of a single copy of $K3$, and so its cohomology
is again in correspondence with that of $K3$.  These states
we can represent by
$$\alpha^i_{-2}|0\rangle$$
where we have introduced new oscillators to keep track
of the fact that this comes from {\it two} copies of $K3$
on top of each other.  We thus see that for the $n=2$ case
we get exactly the same number of states as the physical
states of bosonic oscillators!  Continuing in this way
we find that for ${\cal M}_{1,n}$ we find as many cohomology
elements as the $n$-th level degeneracy of bosonic oscillator
fock space!  This sounds very remarkable!  We are seeing a hint
that somehow bosonic oscillators should be related to type IIA
strings on $K3$.  We will have more to say about this
when we come to a discussion of string dualities in the next section.

Now let us consider further compactification on a circle
down to 5 dimensions.  In this case we can dualize on the
radius of circle and we obtain an equivalent type IIB
compactification on $K3\times S^1$.  The configuration
of D2-branes we have been considering now get mapped
to D3-branes wrapped around cycles of $K3$ and the circle.
If we consider the limit where the size of $K3$ is small,
the effective theory on the $D3$ brane is equivalent to a $1+1$
dimensional sigma model where the target space is ${\cal M}_{1,n}$,
where the spatial direction of the $1+1$ dimensional theory is the
$S^1$.

In the effective theory of type IIB compactified on $K3\times S^1$,
there are various kinds of gauge fields and charges.  The RR
gauge field is one source which couples to D-brane charges.
We can also consider the Kaluza-Klein (KK) $U(1)$ field
corresponding to the gauge field $A_\mu =g_{\mu \theta}$ where
$\theta$ is the circle direction.  We can consider
BPS states with definite D-brane charge and $U(1)$ charge
under the KK $U(1)$.  We can also consider extremal
black holes which carry KK charge as well as D-brane
charge.  It is natural to ask if the Bekenstein-Hawking
entropy of the black hole which is 1/4 of the area (in this
case volume) of the horizon, is valid.  A direct
computation of the black hole metric shows that the
expected entropy is
\eqn\mac{S_{macroscopic}=2\pi \sqrt {Q_H n}}
where $Q_H$ denotes the KK charge and $n$ is related
to the same D-brane charge we have been considering.
To find the analog of these states in the D-brane worldvolume
we simply have to consider states with a definite momentum
along the circle direction (Note that the $U(1)$ gauge symmetry
is the same as translation along the circle).
This implies that we should consider states in the effective
$1+1$ dimensional theory where $L_0-{\overline L}_0=Q_H$.
To preserve BPS condition we need to consider states excited
only on left-movers, i.e. $L_0\not= 0, {\overline L}_0=0$.
For large enough $Q_H$ the number of such states
is given by the asymptotic growth of string degeneracy
for a supersymmetric sigma model on a manifold of real dimension
$4n$ (which is the dimension of ${\cal M}_{1,n}$).  This gives
the central charge $6n$ (taking account of fermions) and
the asymptotic growth is thus
$$S_{microscopic}=2\pi \sqrt{cN\over 6}=2 \pi \sqrt{Q_H n}$$
in agreement with the macroscopic entropy \mac .
This kind of match between black hole entropy
and string state entropy can be generalized to many similar
cases.  A particularly easy case is type II compactification
on tori where the constructions are particularly simple.

\newsec{Web of String Dualities}

We have already mentioned that T-duality connects
certain string theories together:  Type IIA and B get
exchanged under T-duality.  So does heterotic string $E_8\times
E_8$ and $SO(32)$ heterotic string, which get exchanged under
T-duality.  Thus up to compactification and the use of T-duality,
which can be understood using string perturbation theory, we only
have 3 inequivalent string theories: Type II, Type I and heterotic
strings.
It is natural to ask if these three are also connected in some way
with each other.
If there are any connections they should be realized in a non-perturbative
way, because perturbative symmetries have already been accounted for
by T-duality.

\subsec{Type I/heterotic duality}

Let us first consider a possible relation between type I string and
heterotic strings.  In 10 dimensions, the $N=1$ theory with $SO(32)$
gauge symmetry  has two different realizations as a string theory:
Type I strings, as well as $SO(32)$ heterotic strings.  As already
discussed these theories both have the same matter content.  However
they cannot be trivially related, because they have completely different
perturbative properties.  There is one scalar in the theory whose expectation
value plays the role of the coupling constant.  It is thus natural to
speculate that the strong coupling limit of one theory is equivalent to the
weak coupling limit of the other.   There is no way that we know
at the present to prove such a conjecture, because it necessarily
means going beyond perturbative definition of the theory and at the
present we do not have a good understanding of how to do this.
However there are some consequences of this duality
that can in particular be checked.

One particularly nice check is to see how we can
identify the heterotic string as a soliton in the type I string.\foot{
The inverse question is more difficult because type I string does not
couple to a conserved charge as it has both closed and open string sectors.}
Type I has an antisymmetric 2-form gauge field coming from the RR sector.
A D1 brane is a source for this field.  It is natural to ask if
the fields propagating on this string are in any way similar to that of
heterotic strings.  We consider an infinitely long D1 brane.  This
introduces 2 new sectors in the type I theory:  D1-D1 open strings
and D1-D9 open strings (recall that the 32 Chan-Paton factors
can be viewed as having 32 D9 branes).  From the D1-D1 sector
we obtain (after orientifold projection) 8 left and right-moving
bosonic degrees of freedom as well as 8 right-moving fermionic degrees
of freedom.  From the D1-D9 sector we obtain 32 left-moving fermions.
These are precisely the worldsheet degrees of freedom for the SO(32) heterotic
strings in its fermionic formulation.

\subsec{Type II/Type II U-duality and M-theory}
As already remarked in connection with toroidal compactification
of type II strings given that the space of expectation values
for the scalars is of the form  of a coset space $G/H$, and
T-duality is a discrete subgroup of $G$
giving further identification of this space.  This suggests that there
may be further identifications of this space which have non-perturbative
origin.  It is in fact natural to expect that the identification
space is maximal, i.e. $G(Z)$, an integral version of the group $G$,
leading to the moduli space $G/H\times G(Z)$.  The group $G(Z)$
is called the U-duality group.

There are two distinct ways this can be motivated further:
One is to assume an eleven dimensional $N=1$ theory exists.
If so, as mentioned before, toroidal compactifications on $T^{d+1}$
which can also be obtained by type II string compactification on $T^d$ will
give both an $SL(d+1,Z)$ symmetry from the $T^{d+1}$ torus and an
$SO(d,d)$ symmetry from the type II string T-duality.  These groups
do not commute and lead to the group $E_d(Z)$ as discussed before.
This identification in particular suggest that type IIA string which
is a non-chiral $N=2$ theory in $d=10$ should be identified
with a circle compactification of an 11-dimensional $N=1$ theory,
which is called the M-theory.

In 11 dimensional $N=1$ theory the bosonic fields are very simple:
In addition to the metric there is a three index antisymmetric field
$C_{\mu\nu\rho}$ which couples to a 2 dimensional membrane.  Upon
compactification on a circle, we obtain in addition to the
metric and the three index field in 10 dimension,
 one scalar corresponding to the
size of the circle, one vector corresponding to the off-diagonal
components of the metric and an anti-symmetric two form
by considering one index of $C$ to be along the circle.
These are, as expected, the bosonic fields of type IIA strings.
Note in particular that the coupling constant $\lambda$ of type IIA gets
identified with the radius $R$ of the circle.  In particular the precise
relation (which can be easily obtained by
relating the eleven dimensional fields to
ten dimensional fields appearing in type IIA) in terms of the string coupling
constant turns out to be
$$R^3=\lambda^2$$
and thus the eleven dimensional theory $R\rightarrow \infty$ should
emerge in the strong coupling regime $\lambda \rightarrow \infty$.
As a further evidence for the existence of M-theory
and its equivalence to type IIA, one can
ask what would the Kaluza-Klein modes along the circle
be related to in the 10 dimensional theory.  Given the
relation between the fields in M-theory and type IIA strings
these
would be the states which are charged under the vector
field 1-form coming from the RR sector. They couple to D0-branes.
We are thus naturally led to view the D0-branes
(and their bound states) as the KK states
of M-theory.

There is another way to motivate the U-duality group.
Consider type IIB in 10 dimensions.  For this theory
$G=SL(2)$.  In this case the coupling constant $\lambda$ and
the scalar in the RR sector $\chi$ combine to form
a complex field $\tau={i\over \lambda}+\chi$, which transforms
according to the Mobius transformations under $SL(2,Z)$.
The symmetry $\tau\rightarrow \tau+1$ is the statement
that the RR scalar $\chi$ is a periodic field--this
follows from the existence of the magnetic dual
D7-branes.  The other conjectured symmetry
$\tau \rightarrow -1/\tau$ is much more interesting and involves
a strong/weak coupling duality.
This transformation exchanges the two antisymmetric 2-forms
of the NS-NS sector and the RR sector.  Given that
one couples to the fundamental string and the other
to the D1-string, it implies that at strong couplings
the D1-strings should behave the same way as the fundamental string
does at weak coupling. Similarly the D5-brane gets exchanged
with the NS 5-brane.
 As a check on this note that we already had seen
that when $k$ NS 5-branes of type IIB coincide
we obtain a $U(k)$ symmetry by its T-duality
with type IIA on $A_{k-1}$ ALE space.
It turns out that in this strong/weak duality exchange the
D3-brane gets mapped to itself.  If we consider $N$ parallel
D3-branes we get $U(N)$ gauge symmetry and the invariance
of D3-brane under strong/weak duality gets mapped to the conjectured
Olive/Montonen duality on the worldvolume of D3-brane, which
is an $N=4$ theory in $d=4$.
As another check
we can ask whether the degrees of freedom on the D1-string look
like that of the fundamental type IIB strings.
If we consider an infinitely extended D1 string we have
one extra open string D1-D1 sector.  This gives rise
to 8 left- and 8 right-moving bosons and fermions on the
worldsheet, the same degrees of freedom as expected for
type IIB strings.
This is a strong evidence for the existence of
$SL(2,{\bf Z})$ symmetry for 10 dimensional type IIB.

If we compactify type IIB on a circle the T-duality
relates it to type IIA on an inverse sized circle.  The
$SL(2,{\bf Z})$ symmetry of type IIB will be
the U-duality group for type IIA on a circle.  This is
the same as the symmetry expected from M-theory compactification
on $T^2$, for which the $SL(2,{\bf Z})$ symmetry is just the
global reparametrization of the torus (choice of two basis
vectors for the lattice defining the torus). Upon further compactification
the U-duality works as explained above.  Thus the U-duality
symmetry is a consequence of type IIB strong/weak duality
and T-duality of type II strings.

\subsec{$E_8\times E_8$ and M-theory orbifold}
If we consider the 5 string theories in 10 dimensions,
we can ask what is the limit of strong coupling in each case.
We have answered this question for four cases:  Type IIB is
self-dual, type I $SO(32)$ and heterotic $SO(32)$ are dual
to one another, type IIA grows an extra dimension
and becomes the 11-dimensional M-theory.  What about $E_8\times
E_8$?  It is easy to argue that it cannot be self-dual
(by a simple field redefinition involving strong/weak exchange
the theory does not come back to itself).  So what could its
strong coupling limit look like?  There is another question
which we could ask: If we believe in the existence of M-theory
in 11 dimensions, it is natural to consider its compactification
on a circle modded out by a $Z_2$ reflection.  It is easy
to see that this preserves half the supersymmetry and
we end up in 10 dimensions with a theory with $N=1$ supersymmetry.
It is tempting to identify this as the strong coupling limit of
$E_8\times E_8$ heterotic string.
Note that we also have two special fixed points on the circle.
This conjecture is consistent with the presence of these fixed points
(which fill the 10 dimensional
spacetime) each of which is identified
 with a 9-brane carrying the $E_8$ gauge symmetry.
So the fact that we have two factors in this heterotic theory
would be related to the fact that we have two fixed points on a circle
modded out by reflection!  There are various hints that this is the right
picture.

\subsec{Evidence for F-theory}
As we have seen
Type IIA and heterotic $E_8\times E_8$ in ten dimensions
can be viewed as arising from a more fundamental theory
in 11 dimensions, the M-theory.  How about type IIB
and the $N=1$ theory with gauge group $SO(32)$ (which
has realizations as type I or heterotic $SO(32)$ theory)?
In addition
the fact that type IIB has an $SL(2,{\bf Z})$ symmetry
in ten dimensions, begs for a geometric explanation.  From
the M-theory viewpoint if we go to 9 dimensions as explained
above we can see the $SL(2,{\bf Z})$ symmetry as a geometric
symmetry of a torus. But in order to go to the 10 dimensional
limit we have to take the limit of zero size torus, which
gets us away from the domain of validity of geometric
description of M-theory.  Similar
statement applies to the $N=1$ theory with $SO(32)$ gauge group
and how it will have an M-theory realization.
Given the $SL(2,{\bf Z})$ symmetry of type IIB in ten dimensions,
it is natural to postulate the existence of a 12 dimensional
theory, F-theory, which upon compactification on $T^2$, with modulus
$\tau$, would
give rise to type IIB in 10 dimensions, or upon compactification
on a cylinder ($Z_2$ orbifold of $T^2$) would give
$SO(32)$ theory.  Note that then not only the strong weak duality
of type IIB would be manifest, but also the fact that a finite cylinder
has two inequivalent limits gives rise to two theories that
are connected, thus explaining type I/heterotic duality.

This picture is rather suggestive, but various obstacles
would have to be overcome.  For one thing, there is no known
covariant 12 dimensional supergravity theory (and there
are various no-go theorems).  There are further hints
in connection with integrable structures with signature
$(2,2)$ that
a (10,2) signature for F-theory is rather natural to expect.
Perhaps the most reasonable thing to expect for F-theory is a partially
twisted topological theory in 12 dimensions whose
BRST invariant states are the 10 dimensional degrees
of freedom of type IIB strings.

Note that if we compactify type IIB on $S^1$, it should be equivalent
to M-theory on $T^2$.  Moreover type IIB itself should be
identified with F-theory on $T^2$.  Thus we learn that
F-theory on $T^2\times S^1$ is the same as M-theory on $T^2$.

\subsec{String/String Duality}
So far we have seen how the $N=2$ theories, i.e. type II
strings are related and behave for strong couplings.  We have
also seen how $N=1$ theories are related and behave
at strong couplings.  In particular by the time we come
to 9 dimensions there is only one $N=2$ and one $N=1$
theory (up to variation
of moduli).  It is natural to relate also the $N=2$ theories
to $N=1$ theories.  In order to do this we will have
to cut down the number of supersymmetries.  The most natural
way to do this in string theory is to go down to 6 dimensions
and consider type IIA strings compactified on $K3$ and compare
that with $N=1$ strings (heterotic or type I) on $T^4$.
In this case both theories in 6 dimensions will have $N=2$
(non-chiral) supersymmetry.  Moreover the moduli of scalars
for both theories is
$${SO(20,4)\over SO(20)\times SO(4)\times SO(20,4;Z)}\times {\bf R}^+$$
where $R^+$ denotes the coupling constant.  It is natural
to conjecture that the strong coupling limit of one is equivalent
to the weak coupling limit of the other\foot{Note that the notion
of strong coupling limit is a dimension dependent statement because
the volume of the internal compactification rescales the value
of the effective coupling in the lower dimension by ${1\over \lambda^2}
\rightarrow {V\over \lambda^2}$.}.  We have already seen a number
of evidence for this statement.  In particular we have seen that
when $K3$ develops ADE singularity we obtain a non-perturbative
enhanced gauge symmetry
on the type IIA side
due to wrapped D2 branes. This is also matched with a perturbative
enhancement of gauge symmetry on the heterotic side.  We identify
$(P_L,P_R)$ Narain momenta of the heterotic string with a lattice
element of the homology of $K3$.  Having a vanishing 2-cycle
for $K3$ is the same statement as $P_R=0,P_L^2=2$ vector
on the homology lattice of $K3$.  We can also now explain
why in computing the BPS states of type IIA strings
on $K3$ we encountered the partition function of
bosonic string:  The same BPS states on the heterotic side
would map to perturbative BPS states which are elementary
string excitations which on the right-movers are on the ground
state (i.e. no oscillators excited) and on the left-side
are the excited oscillator states of the bosonic string, thus
explaining the mysterious appearance of the partition function
of bosonic strings in the degeneracy of BPS states for type IIA
on K3 that we encountered.  This is the strongest evidence for this
duality (which is sometimes known as the string/string duality).

We can push this duality up in dimension and connect it with
the other dualities already discussed:  First note that type IIA
on $K3$ is by definition M-theory on $K3\times S^1$.
This being equivalent to heterotic on $T^4$ implies, upon
deletion of an $S^1$ from both sides, that M-theory
on $K3$ is equivalent to heterotic on $T^3$.  One evidence
in favour of this is that they both have the same
duality symmetry $SO(19,3)$, which in the case of M-theory
is the symmetry of moduli of Ricci-flat metrics on $K3$.
This gets identified with the heterotic T-duality
upon compactification on $T^3$.  Moreover the overall
size of $K3$ gets identified with the inverse of coupling
of the heterotic theory.

This can be pushed up one more step as follows:
Consider M-theory on elliptic $K3$'s, which we have
already discussed in detail.  Now consider the limit in which
the size of the $T^2$ fiber goes to zero.  Note that M-theory
on $T^2$ in the limit of zero size is identified with type IIB
in 10 dimensions.  Thus we should obtain in this limit
a compactification of type IIB on a sphere times
a large radius $S^1$.  The
sphere is the base
of the elliptic fibration, where the string coupling
constant $\tau$ undergoes $SL(2,{\bf Z})$ monodromies.
Or put differently we can view it as F-theory on elliptic $K3$.
The T-duality group for elliptic $K3$ is $SO(18,2)$ and this
again matches the T-duality group of heterotic string on $T^2$
(which is what one is led to again by deletion of the extra
circle).  At this stage we can also connect this
duality to the dualities already discussed in 10 dimensions:
Consider the limit of $K3=T^2\times T^2/Z_2$,
which is elliptic, in the context of F-theory compactification.  Note that
now we can take the fiber $T^2$ to correspond to a fixed complex structure
with a large value of $\tau$. Given the interpretation of $\tau$ as
the coupling constant of type IIB this point should
be describable in terms of a perturbative type IIB compactification
on $T^2\over Z_2$.  For this to preserve half the supersymmetry,
the $Z_2$ must also exchange the left- and right- movers of strings.
In other words in this limit we are led to the duality between
an orientifold of type IIB on $T^2$ and the heterotic strings
on $T^2$.  If we take the size of the $T^2$ on type IIB side
to be small we would be led back to type I theory in 10 dimensions
compactified on $T^2$, being equivalent to the heterotic
strings on $T^2$.  By elimination of the extra $T^2$ we are
thus led to heterotic/type I duality in 10 dimensions.
Thus heterotic $(T^4)$/type IIA ($K3$) duality is a consequence
of type I/heterotic duality in 10 dimensions.  Of course
we could logically reverse the arrows and say that type I/heterotic
duality is a consequence of type IIA/heterotic duality
in 6-dimensions.

\subsec{The Adiabatic Principle and the web of String Dualities}

In the preceding subsection we have seen the power of using
fiberwise dualities to derive new dualities.  This means
that we start with two theories which are dual
to one another.  If we now consider varying the parameters
of these two theories over a base space then,
at least if we vary the parameters slowly, we expect
the duality to continue to hold on the lower dimensional space.
It turns out that this is a powerful principle to connect
various string dualities in different dimensions
to one another.  It seems to be applicable
{\it even if the adiabaticity assumption is not strictly
valid} and the parameters vary rapidly in some regions
over the base.  It would be very important to understand
why this principle works!
Note that whenever we get some
new physics in lower dimension using this principle which
does not follow in a trivial way from the dimensional
reduction of the higher dualities, the new physics is hidden at
 the points where the adiabaticity assumption fails.
At any rate, taking the adiabaticity principle for
granted we can now generate new dualities and new
physics in lower
dimensions.

Let us just give one example of the application of this principle:
Consider F-theory on $K3$ duality with heterotic on $T^2$.  Now
we can fiber both of these spaces over a ${\bf P}^1$, in such
a way that on the heterotic side we get a $K3$ compactification
of heterotic strings.  On the F-theory side we get a Calabi-Yau
three-fold which is an elliptic fibration over a base space
which is itself a ${\bf P}^1$ fiber space over ${\bf P^1}$.
Given a base space the elliptic Calabi-Yau is fixed (up
to variation of moduli).  Thus the number of inequivalent
such choices is related to the number of inequivalent
${\bf P}^1$ fiber spaces over ${\bf P}^1$.  These
are parameterized by an integer $n$, and are known
as the Hirzebruch surface $F_n$.  The case
of the elliptic CY 3-fold over $F_0=\bf P^1\times \bf P^1$
is the one we have already discussed in detail in section 2.
For $n\leq 12$ there does exist a smooth Calabi-Yau.  On the
heterotic side it turns out that this $n$ is mapped to the
choice of the splitting of the instanton number between the
two $E_8$'s.  Recall that we need the total number of
instantons to be $24$. The two instanton numbers turn out
to be $12+n,12-n$.  If we further compactify on $S^1$
we find duality between M-theory on the corresponding
Calabi-Yau threefold with heterotic strings on $K3\times S^1$.
Continuing one more step we find duality between type IIA on
Calabi-Yau threefold with heterotic strings on $K3\times T^2$.
These will give $N=2$ theories in four dimensions.
For example we learn that type IIA compactification
on elliptic Calabi-Yau over $F_0$ is equivalent to
$E_8\times E_8$ heterotic
strings on $K3\times T^2$ with instanton numbers 12 for each $E_8$.
It turns
out that the moduli space of the vector multiplet
on the type IIA side does not receive quantum string corrections
(since the string coupling is in a hypermultiplet) and thus
exact non-trivial results for the moduli of $N=2$ field theories can
be obtained from classical string considerations on type IIA side.
Moreover the moduli of hypermultiplets on the heterotic side
does not receive quantum corrections because the heterotic string
coupling constant is in a vector multiplet.  Thus the classical
moduli of hypermultiplets on the heterotic side gives the exact
quantum corrected result for the hypermultiplet moduli of type IIA.
This phenomenon is also known as the `second quantized mirror symmetry'.

We can also consider the fibration of the 8 dimensional
heterotic/F-theory on a two complex dimensional base, in which
case we obtain duality between F-theory on Calabi-Yau fourfold
with heterotic strings on Calabi-Yau threefold.  This is a duality
between two theories with $N=1$ in $d=4$.

\newsec{Speculations and Open Questions}
In this paper we have reviewed some aspects of
string theories with emphasis on their solitonic/non-perturbative
aspects.    We have also seen that duality of one string theory
with another can occur in a rather non-trivial way.  Due
to accumulation of massive evidence in their favor
these dualities have the status of conjectural facts!
These dualities exchange what looks like classical results
in one theory to highly non-trivial quantum results on the other
side.  It clearly demonstrates the fact that what constitutes
quantum corrections is partly a matter of convention.
This suggests a rather revolutionary reformulation of quantum mechanics.
This is a natural extension of various relativity principles
we have learned in physics in this century:
In general theory of relativity we have learned
that many physically reasonable statements make sense
only when we specify relative to which observers
those statements are made.  In quantum mechanics
we learn that the very notion of reality is relative
to an experimental setup.  Here we are learning that the notion
of quantum versus classical is also relative to which
theory we measure it from.

One basic fact we have seen
many times in the context of string duality is that in each regime
of parameter space there is at most one simple description of the
physics.  This is somewhat similar to the complementarity principle
of Bohr in the context of quantum mechanics:  His view was that
depending on the experiment one should either use the wave
picture or the particle picture but not both at the same time.\foot{We know
that quantum mechanics ultimately developed in a
direction somewhat different from this view of Bohr, in that
the wave description is always the correct description.
In the case of string dualities we have strong evidence
that none of the string theories (nor M or F theory for that matter)
 are supreme to any other and each one is as fundamental as any other.
This suggests that perhaps Bohr's original view
of complementarity may have more merits than the present view
of quantum mechanics leads us to believe.}

Despite the tremendous progress made recently, we should
not lose sight of the fact that the most interesting physical
questions are still unanswered by these developments.  In particular
a pessimist might say that the recent discovery of the
non-perturbative consistency
of string theories in various dimensions with varying number
of supersymmetries inevitably leads to the conclusion that the universe
we live in is not apparently selected by consistency conditions
alone!  This may also make the precise realization of our
universe in the context of string theory more difficult.
In fact since we do not have a single example
of a non-supersymmetric string vacuum which we can claim
to understand deeply enough, we are still rather far from the ultimate
goal of string theory.  This is also connected with the
fact that non-supersymmetric theories will have
no known criterion to rule out a non-zero cosmological
constant, in apparent contradiction with the observed universe.
Clearly we have still a lot to learn.

Another aspect of non-supersymmetric situations which
we do not know much about involve aspects of black hole
far away from being extremal (i.e. far from being BPS).
In these situations we run into the usual puzzles
of black holes (the information puzzle in particular).
It seems natural to speculate that the two puzzles
of cosmological constant and black hole information puzzle
are related to one another, and what resolves one
will also resolve the other issue.  At any rate
we have not made any real progress on either of these fronts.

We have learned a great deal about non-perturbative aspects of string theory.
One of the main reasons we believe in duality is that string theory
is a concrete theory with concrete properties and any statement
about duality will have many consequences which can be readily
checked using tools available in string theory.   The same
cannot be said yet, unfortunately, about M-theory nor F-theory.
I thus believe that until we
have a deeper understanding of these non-perturbative
aspects of string theory and what is the right way to think
about them, it would be prudent
to continue calling the general subject we are studying
by its own name:  `string theory'!

We would like to thank B. Zwiebach for suggestions for improving
the presentation.

This research is supported in part by NSF grant  PHY-92-18167.
\listrefs
\end